%% file: jsac-mmwave.tex
\documentclass[journal]{IEEEtran}

\IEEEoverridecommandlockouts

\makeatletter
\def\markboth#1#2{\def\leftmark{\@IEEEcompsoconly{\sffamily}\MakeUppercase{\protect#1}}%
\def\rightmark{\@IEEEcompsoconly{\sffamily}\MakeUppercase{\protect#2}}}
\makeatother

\usepackage[english]{babel}
\usepackage{color}
\usepackage{lipsum}
\usepackage{cite}
\usepackage{graphicx}
\usepackage{amsmath}
\usepackage{amsfonts}
\usepackage{array}
\usepackage{verbatim}
\usepackage{listings}
\usepackage{url}
\usepackage{enumerate}
\usepackage{multirow}

\usepackage{epsfig}
\usepackage{multicol}
\usepackage[font=footnotesize]{caption}
\usepackage[font=scriptsize]{subcaption}
\usepackage{tikz}
\usepackage{pgfplots}
\pgfplotsset{compat=newest} 
\pgfplotsset{plot coordinates/math parser=false} 
\newlength\fheight
\newlength\fwidth
\usetikzlibrary{patterns,decorations.pathreplacing,backgrounds,calc,arrows,arrows.meta}
\definecolor{SchoolColor}{RGB}{0.71, 0, 0.106}
\definecolor{chaptergrey}{rgb}{0.61, 0, 0.09} 
\definecolor{midgrey}{rgb}{0.4, 0.4, 0.4}
\definecolor{chaptergreen}{rgb}{0.09, 0.612, 0}
\definecolor{chapterpurple}{rgb}{0.522, 0, 0.612}
\definecolor{chapterlightgreen}{rgb}{0, 0.612, 0.522}

\usepackage{algorithm}
\usepackage[noend]{algpseudocode}

\usepackage{lscape}
\usepackage{setspace}

\usepackage{etoolbox}

\addto\captionsenglish{}

\renewcommand{\arraystretch}{2}

\newcommand{\bi}{\begin{itemize}}
\newcommand{\ei}{\end{itemize}}
\newcommand{\be}{\begin{equation}} 
\newcommand{\ee}{\end{equation}}

\def\beq{\begin{equation}}
\def\eeq{\end{equation}}
\def\beqa{\begin{eqnarray}}
\def\eeqa{\end{eqnarray}}
\def\beqan{\begin{eqnarray*}}
\def\eeqan{\end{eqnarray*}}

\usepackage{threeparttable}
\usepackage{tabularx}
   \usepackage{multirow}
   \usepackage{booktabs}

   \usepackage{array, blindtext}
   \usepackage{wrapfig}
\usepackage{pdfpages}

\newif\iftikz
\tikztrue

\graphicspath{{./figures/}}
\title{Improved Handover Through Dual Connectivity \\ in 5G mmWave Mobile Networks}
\author{{{Michele Polese},~\IEEEmembership{Student Member, IEEE}, {Marco Giordani},~\IEEEmembership{Student Member, IEEE},\\{Marco Mezzavilla},~\IEEEmembership{Member, IEEE},
 {Sundeep Rangan},~\IEEEmembership{Fellow, IEEE}, {Michele Zorzi},~\IEEEmembership{Fellow, IEEE}}

\thanks{Michele Polese, Marco Giordani and Michele Zorzi are with the Department of Information Engineering, University of Padova, Padova, Italy (email: polesemi@dei.unipd.it; giordani@dei.unipd.it; zorzi@dei.unipd.it).}
\thanks{Marco Mezzavilla and Sundeep Rangan are with NYU WIRELESS, Tandon School of Engineering, New York University, Brooklyn, NY (email: mezzavilla@nyu.edu; srangan@nyu.edu)} 
}

\makeatletter
\patchcmd{\@maketitle}
  {\addvspace{0.5\baselineskip}\egroup}
  {\addvspace{-1\baselineskip}\egroup}
  {}
  {}
\makeatother

\begin{document}
\maketitle


\tikzstyle{startstop} = [rectangle, rounded corners, minimum width=2cm, minimum height=0.5cm,text centered, draw=black]
\tikzstyle{io} = [trapezium, trapezium left angle=70, trapezium right angle=110, minimum width=3cm, minimum height=1cm, text centered, draw=black]
\tikzstyle{process} = [rectangle, minimum width=2cm, minimum height=0.5cm, text centered, draw=black, align=center]
\tikzstyle{decision} = [ellipse, minimum width=2cm, minimum height=1cm, text centered, draw=black]
\tikzstyle{arrow} = [thick,<->,>=stealth]
\tikzstyle{line} = [thick,>=stealth]
\tikzstyle{darrow} = [thick,<->,>=stealth,dashed]
\tikzstyle{sarrow} = [thick,->,>=stealth]

\begin{abstract}
The millimeter wave (mmWave) bands offer the possibility of orders of magnitude
greater throughput for fifth generation (5G) cellular systems.
However, since mmWave signals are highly susceptible to blockage, channel quality
on any one mmWave link can be extremely intermittent.  This paper implements
a novel dual connectivity protocol that enables mobile user equipment (UE) devices
to maintain physical layer connections to 4G and 5G  cells simultaneously.
A novel uplink control signaling system combined with a local coordinator enables
rapid path switching in the event of failures on any one link.  
This paper 
provides the first comprehensive end-to-end evaluation of handover mechanisms in mmWave cellular systems.
The simulation framework includes 
detailed measurement-based channel models to realistically capture
spatial dynamics of blocking events, as well as the full details of MAC, RLC and  transport protocols. Compared to conventional handover mechanisms, the study reveals significant
benefits of the proposed method under several metrics.
\end{abstract}

\begin{picture}(10,-300)(10,-330)
\put(0,0){
\put(0,0){\scriptsize This paper was accepted for publication on the IEEE JSAC Special Issue on} 
\put(0,-10){\scriptsize Millimeter Wave Communications for Future Mobile Networks.}
\put(0,-20){\scriptsize DOI 10.1109/JSAC.2017.2720338}}
\end{picture}


\section{Introduction}

The millimeter wave (mmWave) bands -- roughly corresponding to
frequencies above 10~GHz~--~have attracted considerable attention for
next-generation cellular wireless systems
~\cite{KhanPi:11-CommMag,rappaportmillimeter,RanRapE:14,andrews2014will,ghosh2014millimeter}.
These bands offer orders of magnitude more spectrum than conventional cellular
frequencies below 3 GHz
-- up to 200 times by some estimates \cite{KhanPi:11-CommMag}.
However, a key challenge in delivering robust service in the mmWave
bands is channel intermittency:
MmWave signals are completely blocked by many common
building materials such as brick and mortar,
\cite{Allen:94,singh2007millimeter,KhanPi:11-CommMag,rappaport_channel_model},
and even the human body can cause
up to 35~dB of attenuation \cite{lu2012modeling}.
As a result, UE mobility, combined with small movements of obstacles and reflectors,
or even changes in the orientation of a handset relative to the body or a hand,
can cause the channel to rapidly appear or disappear.

One of the main tools to improve the robustness of mmWave systems is
\emph{multi-connectivity} \cite{ghosh2014millimeter}:
Each mobile device (UE or user equipment in 3GPP terminology) maintains
connections to multiple cells, possibly including both 5G mmWave cells and/or
conventional 4G cells.
In the event that one link is blocked, the UE can find alternate routes to
preserve the connection.  In cellular systems, this robustness is called \emph{macro-diversity}
and is particularly vital for mmWave systems \cite{ghosh2014millimeter}.

How to implement multi-connectivity in the network layer for mmWave systems
remains largely an open problem.  Current 3GPP cellular systems offer multiple mechanisms
for fast switching of paths between different cells including
conventional handover, multi-connectivity and carrier aggregation~-- these
methods are summarized below.  However, mmWave systems present unique challenges:
\begin{itemize}
\item Most importantly, the  dynamics of mmWave channels imply
that the links to any one cell can  deteriorate rapidly, necessitating much faster link detection
and re-routing~\cite{RanRapE:14}.
\item Due to the high isotropic pathloss, mmWave signals are transmitted in narrow beams,
typically formed with high-dimensional phased arrays.
In any link, channel quality must be continuously scanned across multiple possible
directions which can dramatically increase the time it takes to detect that the link has failed
and a path switch is necessary \cite{MedHoc2016,MCJournal2016}.
\item One of the main goals of 5G is to achieve ultra-low latency~\cite{Boccardi} (possibly $< 1$ ms).  Thus, service unavailability
during path switches must be kept to a minimum.
\end{itemize}

\subsection{Contributions}
To address these challenges, in this paper we expand the main results and findings of our previous works \cite{MedHoc2016,MCJournal2016,EW2016,simutoolspolese}, to provide the first global end-to-end comprehensive evaluation of handover and path switching of mmWave systems under realistic dynamic scenarios, and assess how a dual-connectivity\footnote{Although many of the ideas and techniques discussed in this paper apply to more general multi-connectivity scenarios, for concreteness in the following we will specifically refer to \emph{dual} connectivity, in which a UE is simultaneously connected to one 5G mmWave base station and one legacy LTE eNB.} (DC) approach can enable faster, more robust and better performing mobility management schemes.

In particular, in \cite{MedHoc2016} and \cite{MCJournal2016}  we proposed a novel multi-connectivity uplink measurement framework that, with the joint effort of the legacy LTE frequencies,  enables fair and robust cell selection, in addition to  efficient and periodic tracking of the user, suitable for several control-plane cellular applications (i.e., we showed that periodic measurement reports can be used to trigger handovers or adapt the beams of the user and its serving cell, to grant good average throughput and deal with the channel dynamics experienced at mmWave frequencies).
In \cite{EW2016}, we evaluated the tracking performance of a user's signal quality considering real experiments in common blockage scenarios, combined with outdoor statistical models.
Finally, in \cite{simutoolspolese},  we discussed two possible ways to integrate 5G and LTE networks to improve the reliability of next-generation mobile users, and described a preliminary ns--3 simulation framework to evaluate the performance of both.

By extending our previous contributions, in this paper we also propose:
\begin{itemize}
\item The use of a DC scheme to enable the base stations to efficiently track the UE
channel quality along multiple links and spatial directions within those links.
In addition, to allow fast detection of link failures, 
we demonstrate that the uplink control signaling
enables the network to track the angular
directions of communication to the UE on all possible links
simultaneously, so that, when a path switch is necessitated, no directional search needs
to be performed (this approach greatly saves switch time, since directional scanning
dominates the delay in establishing a new link  \cite{Barati,barati2015initial}).

\item The use of a local coordinator that manages the traffic
between the cells.  The coordinator performs both control plane tasks of path switching
and data plane tasks as a traffic anchor, at the Packet Data Convergence Protocol (PDCP) layer.
In conventional cellular systems, these control and data plane functions are
performed in the Mobility Management Entity (MME) and Serving Gateway (S-GW),
which are often far from the cells.  In contrast, the local coordinator
is placed in close proximity to the cells, significantly reducing the path switch time.

\item The design of faster network handover procedures (namely \textit{fast switching} and \textit{secondary cell handover}) that, by exploiting our DC framework, improve the mobility management in mmWave networks, with respect to the standard standalone hard handover (HH) scheme. These procedures are controlled by the LTE Radio Resource Control (RRC) layer and, since the UE is connected to the LTE and the mmWave eNBs, it is possible to perform quick fallback to LTE with the fast switching command.

\item A dynamic time-to-trigger (TTT) adaptation to enhance the switch
decision timing in highly uncertain link states.
\end{itemize}

Moreover, we evaluate the proposed switching and handover protocols
by extending the evaluation methodology we have developed in \cite{mmWaveSim,phy5g,zhang2016transport,simutoolspolese}.
The ns3-based framework we implemented for this work makes it possible to use detailed measurement-based
channel models that can account for both the spatial characteristics of the channel
and the channel dynamics arising from blocking and other large-scale events,
which is important for a detailed and realistic assessment.
In addition, the simulator features a complete MAC layer with HARQ, all the network-layer
signaling, and an end-to-end transport protocol.  
We believe that this is the first exhaustive contribution which provides a global evaluation of the performance of a dual-connectivity architecture with respect to a traditional standalone HH scheme in terms of handover and mobility management specifically tailored to a dynamic mmWave scenario.
In particular, we simulated the user's motion in a typical urban environment. Separately, actual local blockage dynamics were measured and superimposed on the statistical channel model, to obtain a realistic spatial dynamic channel model. We believe that this is the first work in which such detailed mmWave dynamic models have been used in studying handover.

Our study reveals several important findings on the interaction of 
transport layer mechanisms, buffering, and its interaction with physical-layer
link tracking and handover delays. 
 We also demonstrate that the proposed dual connectivity framework offers significant performance improvements in the handover management of an end-to-end network with mmWave access links, including (i)   reduced packet loss, (ii) reduced control signaling, (iii) reduced latency, and (iv) higher throughput stability. 
 Moreover, we show that a dynamic TTT approach should be preferred for  handover management, since it can deliver non-negligible improvements in specific mobility scenarios in which  state-of-the-art methods fail.

\subsection{Related Work}
\label{sec:rel_work}

Dual connectivity to different types of cells (e.g., macro and pico cells) has been proposed in Release $12$ of Long Term Evolution-Advanced (LTE-A) \cite{3GPP_DC} and in  \cite{sota_MC2}.  However, these systems were designed for conventional sub-6 GHz
frequencies, and the directionality and variability of the channels typical of mmWave 
frequencies were not addressed. 
Some other previous works, such as \cite{sota_MC},   
consider only the bands under $6$ GHz for the control channel of 5G networks, to provide robustness against blockage and a wider coverage range, but this solution could not provide the high capacities that can be obtained  when exploiting  mmWave frequencies. 
The potential of combining legacy and mmWave technologies in outdoor scenarios has also been investigated in  \cite{EuCnCBackhaul2016}, highlighting the significant benefits that a mmWave network achieves with flexible, dynamic support
from LTE technologies.
Articles \cite{5Gmulticonnectivity2016,MC} propose a multi-connectivity framework as a solution for mobility-related link failures and throughput degradation of cell-edge users, enabling increased reliability with different levels of mobility.

Although the literature on handover in more traditional sub-6 GHz heterogeneous networks is 
quite mature, papers on handover management for mmWave 5G cellular are very recent, and research in this field has just started.
The survey in \cite{Yan20101848} presents multiple vertical handover decision algorithms that are essential for heterogeneous wireless networks, while article \cite{PappalardoHandover2016} investigates the management of the handover process between macro, femto and pico cells, proposing a theoretical model to characterize the performance of a mobile user in heterogeneous scenarios as a function of various handover parameters. However, these works are focused on low frequency legacy cellular systems.
When dealing with mmWaves, frequent handover, even for fixed UEs, is a potential drawback that needs to be addressed.
In \cite{handoff}, the handover rate in 5G systems is investigated 
 and in \cite{HO_train} a  scheme for handover management in high-speed railway is proposed by employing the received signal quality from measurement reports. 
 In \cite{HO_het,HO_het_2}  the impact of user mobility in multi-tier heterogeneous networks is analyzed and a framework is proposed to solve the dynamic admission and mobile association problem in a wireless system with mobility.
Finally, the authors of \cite{magazineControlPlane} present an architecture for mobility, handover and routing management.

\section{Framework Description For Dual Connectivity}
\label{sec:MCP}

We propose a dual connectivity architecture, introduced here for the control and user planes as an extension of 3GPP's LTE DC proposal~\cite{3GPP_DC} to the needs of mmWave communications. In the proposed solution, the UE is simultaneously connected to both LTE and mmWave eNBs. The LTE cell is a backup for the user plane: since the UE is already connected, when the signal quality of the mmWave link degrades, there is no need to perform a complete handover; a single RRC control message from the LTE eNB to the UE is enough. Moreover, for the control plane, this scheme enables a coordinated measurement collection as described in~\cite{MedHoc2016,MCJournal2016}. 
\begin{figure}[t]
  \centering
  \includegraphics[width=0.85\columnwidth]{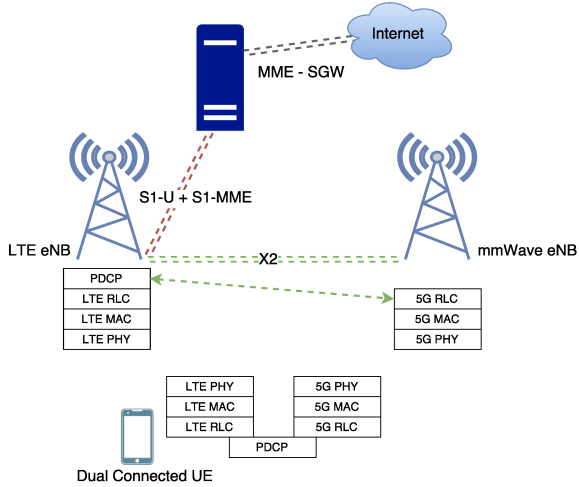}
  \caption{LTE-5G tight integration architecture.}
  \label{fig:arch}
\end{figure}
Fig.~\ref{fig:arch} shows a block diagram of the proposed architecture, as presented in~\cite{simutoolspolese}. For each DC device there is a single connection point to the core network (CN), through the S1 interface that links the LTE eNB to the CN: the mmWave eNB does not exchange control messages with the MME. The two eNBs are connected via an X2 link, which may be a wired or wireless backhaul. Each LTE eNB coordinates a cluster of mmWave eNBs which are located under its coverage. Notice that the coordinator may also be placed in a new node in the core network, or can be based on Network Function Virtualization (NFV) logic.

In the following paragraphs, we will present in detail how the DC framework enables (i) channel monitoring over time, (ii) a PDCP layer integration across different radio access networks, and (iii) faster network handover procedures. 

\subsection{Control Plane For Measurement Collection}
\label{sec:measurement_framework}

Monitoring the channel quality is an essential component of any modern cellular system, since it is the basis for enabling and controlling many network tasks including rate prediction, adaptive modulation and coding, path selection and also handover.
In this work, we follow the multi-cell measurement reporting system proposed in~\cite{MedHoc2016,MCJournal2016},
where each UE
 directionally broadcasts a sounding reference signal (SRS) in a time-varying direction that continuously
 sweeps the angular space. Each potential serving cell scans all its angular directions
 and monitors the strength of the received SRS, building a report table (RT) based on the channel quality of each receiving direction,
to better capture the dynamics of the channel\footnote{Unlike in
traditional LTE systems, the proposed framework is based on
the channel quality of uplink (UL) rather than downlink (DL) signals. This eliminates the need for the UE
to send measurement reports back to the network and thereby removes a possible point of failure in the control
signaling path.}.
A centralized coordinator (which may reside in the LTE eNB)  obtains complete directional knowledge from all the RTs sent by the potential cells
in the network to make the optimal serving cell selection and
scheduling decisions. In particular, due to the knowledge  gathered on the signal quality in each angular direction for each eNB-UE pair, the coordinator is able to match the beams of the transmitter and of the receiver to provide  maximum performance.
 
In this work, we assume that nodes select one of a finite number of directions for measuring the signal quality,
and we let $N_{\rm eNB}$ and $N_{\rm UE}$ be the number of directions at each eNB
and UE, respectively.  
Supposing that $M$ cells  are deployed within the coverage of the coordinator, the procedure works as follows.

\begin{table*}[!t]
\centering
\small
\renewcommand{\arraystretch}{1}
\begin{tabular}{@{}lr@{}}
\toprule
\multicolumn{2}{@{}c}{RT (mmWave eNB$_j$)} \\
\midrule
UE$_1$ & SINR$_{1,j}$ \\
UE$_2$ & SINR$_{2,j}$ \\
\dots &\dots\\
UE$_N$ &  SINR$_{N,1}$\\
\bottomrule
\end{tabular}
\quad\quad\quad
\begin{tabular}{@{}lrrr@{}}
\toprule
\multicolumn{4}{@{}c}{Complete Report Table (CRT)} \\
\midrule
UE & mmWave eNB$_1$ & \dots & mmWave eNB$_M$\\
\midrule
UE$_1$ & SINR$_{1,1}$ &  \dots & SINR$_{1,M}$ \\
UE$_2$ & SINR$_{2,1}$ &  \dots & SINR$_{2,M}$ \\
\dots &\dots  &\dots &\dots \\
UE$_N$ & SINR$_{N,1}$ &  \dots & SINR$_{N,M}$ \\
\bottomrule
\end{tabular}
\caption{An example of RT (left) and CRT (right), referred to $N$ users and $M$ available mmWave eNBs in the network. We suppose that the UE can send the sounding signals through $N_{\rm UE}$ angular directions and each mmWave eNB  can receive them through $N_{\rm eNB}$ angular directions. Each pair is the maximum SINR measured in the best direction between the eNB and the UE.}
\label{tab:RT}
\end{table*}

\subsubsection{First Phase -- Uplink Measurements}
Each UE directionally broadcasts uplink sounding reference signals in dedicated slots, steering through directions $d_1,\dots,d_{N_{\rm UE}}$, one  at a time, to cover the whole angular space.
The SRSs are scrambled by locally unique identifiers (e.g., C-RNTI) that are known to the
mmWave eNBs and can be used for channel estimation.
If analog beamforming is used,
each mmWave eNB  scans through directions $D_1,\dots,D_{N_{\rm eNB}}$ one at a time or, if digital
beamforming is applied, collects measurements from all of them at once.
Each mmWave eNB fills a RT, as in Table \ref{tab:RT} left, whose entries represent the highest  SINR between  UE$_i,\: i=1,\dots,N$, transmitting through its best direction $d_{\rm UE,opt} \in \{d_1,\dots,d_{N_{\rm UE}}\} $, and eNB$_j,\: j=1,\dots,M$, receiving through its best possible direction $D_{\rm eNB,opt} \in \{D_1,\dots,D_{N_{\rm eNB}} \}$:
\beq
\text{SINR}_{i,j} = \max_{\substack{d_{\rm UE}=d_1,\dots,d_{N_{\rm UE}}\\ D_{\rm eNB} =D_1,\dots,D_{N_{\rm eNB}}}}  \text{SINR}_{i,j}(d_{\rm UE},D_{\rm eNB})
\label{eq:max_SINR}
\eeq

\subsubsection{Second Phase -- Coordinator Collection}
Once the RT of each mmWave eNB has been filled for each UE, each mmWave cell sends
this information, through the X2 link, to the coordinator\footnote{The complexity of this framework resides in the central coordinator, which has to aggregate the RT from the $M$ mmWave eNBs that are under its control and  perform for each of the $N$ UEs a search operations among $M$ entries. As the number of the mmWave eNBs $M$ increases, the search space increases linearly.} which, in turn, builds a complete report table (CRT), as depicted in Table \ref{tab:RT} right.
When accessing the CRT, the
optimal mmWave eNB (with its optimal  direction $D_{\rm eNB,opt}$) is selected for each UE (with optimal direction $d_{\rm UE,opt}$), considering the absolute maximum SINR in each CRT's row. The criterion with which the best mmWave eNB is chosen will be described in Section~\ref{sec:handover}.

\subsubsection{Third Phase -- Network Decision}
The coordinator reports  to the UE, on a legacy LTE connection, which
 mmWave eNB  yields the best performance, together with the optimal direction $d_{\rm UE,opt}$ in which  the UE should steer its beam, to reach the candidate serving mmWave eNB in the optimal way. 
The choice of using the LTE control link is motivated by the fact that  
the UE may not be able to receive from the optimal mmWave link if not properly configured and aligned.
Moreover, since path switches and handover events in the mmWave regime
are commonly due to link failures, the control link to the serving mmWave cell may not be available.
Finally, the coordinator also notifies the designated mmWave eNB, through the X2 link, about the optimal direction $D_{\rm eNB,opt}$ in which to steer the beam,  for serving each UE.
We highlight that the procedure
described in this section allows to optimally adapt the beam even when a handover is
not strictly required. In particular, if the user's optimal mmWave eNB is the same as the current one, but a new steering direction pair ($d_{\rm UE,opt},D_{\rm eNB,opt}$)
is able to provide a higher SINR to the user, a beam switch is prompted, to realign with
the eNB and guarantee better communication performance.

According to \cite{CISS}, we assume that the SRSs  are transmitted periodically once every $T_{\rm per}=200$~$ \mu$s seconds, for a duration of $T_{\rm sig}~=~10  \: \mu$s seconds (which is deemed sufficient to allow proper channel estimation at the receiver), to maintain a constant overhead  $\phi_{\rm ov} = T_{\rm sig}/T_{\rm per} = 5 \%$. The switching time for beam switching is in the scale of nanoseconds, and so it can be neglected~\cite{chandra2014adaptive}.
The scanning for  the SRSs for each UE-eNB direction and the filling of each RT  require 
$N_{\rm eNB}N_{\rm UE}/L$ scans, where $L$ is the number of directions in which the
receiver can look at any one time.  Since there is one scanning opportunity every $T_{\rm per}$
seconds, the total delay is
\beq
\label{eq:D}
    	D = \frac{N_{\rm eNB}N_{\rm UE}T_{\rm per}}{L}.
\eeq

The value of $L$ depends on the beamforming (BF) capabilities.  In the uplink-based design, $L=1$ if
the eNB receiver has analog BF and $L=N_{\rm eNB}$ if it has a fully digital transceiver. According to Eq.~\eqref{eq:D}, the value of $D$ is independent of the number of users and of the MAC layer scheduling. Since each UE sends its sounding reference signals at the same time and the mmWave eNBs synchronously receive those messages through exhaustive search schemes, the proposed framework scales well with the network density.

\begin{table}[t!]
\centering
\small
\renewcommand{\arraystretch}{1}
\begin{tabular}{@{}lll@{}}
\toprule
BF Architecture &  & Delay $D$\\
\cline{1-2}\rule{0pt}{3ex}
mmWave eNB Side & UE Side & \\
\midrule
Analog & Analog &  $25.6$ ms  \\
Hybrid & Analog & $25.6/L$ ms \\
Digital & Analog & $1.6$ ms\\
\bottomrule
\end{tabular}
\caption{Delay $D$ for each mmWave eNB to fill each RT. A comparison among different BF architectures (analog, hybrid and fully digital) is reported. We assume $T_{\rm sig} = 10 \: \mu$s, $T_{\rm per} = 200 \: \mu$s (to maintain an overhead $\phi_{\rm ov} = 5\%$), $N_{\rm UE} = 8$ and $N_{\rm eNB} = 16$.}
\label{tab:BF_arch}
\end{table}

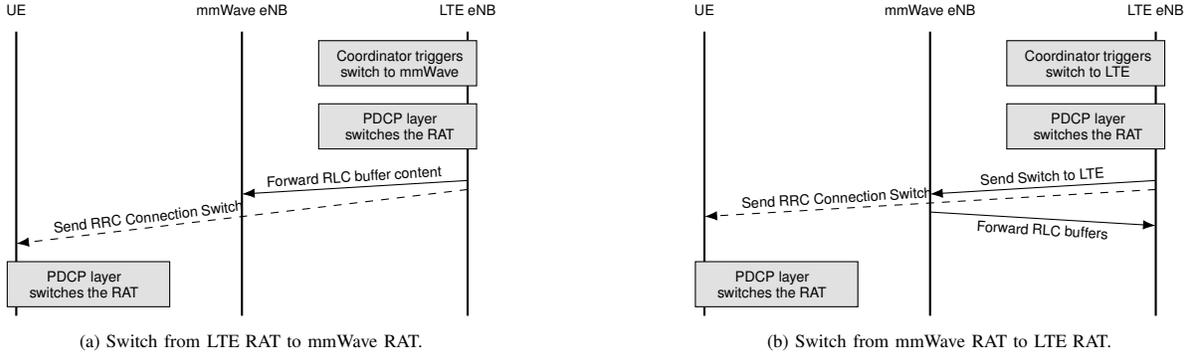
\begin{figure*}[t]
\begin{subfigure}[t]{0.5\textwidth}
\setlength{\belowcaptionskip}{0cm}
\centering
  \begin{tikzpicture}[font=\sffamily\small, scale=0.6, every node/.style={scale=0.6}]
    \coordinate (UeBegin) at (0,10.5);
    \coordinate (UeEnd) at (0,4.2);
    \coordinate (mmWaveEnbBegin) at (5,10.5);
    \coordinate (mmWaveEnbEnd) at (5,4.2);
    \coordinate (lteEnbBegin) at (10,10.5);
    \coordinate (lteEnbEnd) at (10,4.2);
    
    \node [above of=UeBegin, yshift=-15pt] (UElabel) {UE};
    \node [above of=mmWaveEnbBegin,yshift=-15pt] (mmWaveEnbLabel) {mmWave eNB};
    \node [above of=lteEnbBegin,yshift=-15pt] (lteEnbLabel) {LTE eNB};

    \begin{scope}[on background layer]
      \draw [thick, black] (UeBegin) -- (UeEnd);
      \draw [thick, black] (mmWaveEnbBegin) -- (mmWaveEnbEnd);
      \draw [thick, black] (lteEnbBegin) -- (lteEnbEnd);
    \end{scope}

    \filldraw[fill=midgrey!20] (6.7,10.3) rectangle (10.2,9.3);
    \node [align=center] (triggerSwitch) at (8.5,9.8) {Coordinator triggers\\switch to mmWave};

    \filldraw[fill=midgrey!20] (6.7,8.9) rectangle (10.2,7.9);
    \node [align=center] (enbSwitch) at (8.5,8.4) {PDCP layer \\ switches the RAT};
    \draw[-{Latex}] (10, 7.2) -- node[sloped, anchor=center, above] {Forward RLC buffer content} (5, 6.9);

    \draw [-{Latex}, dashed] (10,7.0) -- node[sloped, anchor=center, above] {Send RRC Connection Switch$\quad\quad\quad\quad\quad\quad\quad\quad\quad\quad\quad\quad\quad$} (0, 5.8);
    \filldraw[fill=midgrey!20] (-0.2,5.4) rectangle (3.4,4.4);
    \node [align=center] (ueSwitch) at (1.5,4.9) {PDCP layer \\ switches the RAT};

  \end{tikzpicture}
  \caption{Switch from LTE RAT to mmWave RAT.}
  \label{fig:lteToMm}
\end{subfigure}
\hfill
\begin{subfigure}[t]{0.5\textwidth}
\setlength{\belowcaptionskip}{0cm}
\centering
  \begin{tikzpicture}[font=\sffamily\small, scale=0.6, every node/.style={scale=0.6}]
    \coordinate (UeBegin) at (0,10.5);
    \coordinate (UeEnd) at (0,4.2);
    \coordinate (mmWaveEnbBegin) at (5,10.5);
    \coordinate (mmWaveEnbEnd) at (5,4.2);
    \coordinate (lteEnbBegin) at (10,10.5);
    \coordinate (lteEnbEnd) at (10,4.2);
    
    \node [above of=UeBegin, yshift=-15pt] (UElabel) {UE};
    \node [above of=mmWaveEnbBegin,yshift=-15pt] (mmWaveEnbLabel) {mmWave eNB};
    \node [above of=lteEnbBegin,yshift=-15pt] (lteEnbLabel) {LTE eNB};

    \begin{scope}[on background layer]
      \draw [thick, black] (UeBegin) -- (UeEnd);
      \draw [thick, black] (mmWaveEnbBegin) -- (mmWaveEnbEnd);
      \draw [thick, black] (lteEnbBegin) -- (lteEnbEnd);
    \end{scope}

    \filldraw[fill=midgrey!20] (6.7,10.3) rectangle (10.2,9.3);
    \node [align=center] (triggerSwitch) at (8.5,9.8) {Coordinator triggers\\switch to LTE};

    \filldraw[fill=midgrey!20] (6.7,8.9) rectangle (10.2,7.9);
    \node [align=center] (enbSwitch) at (8.5,8.4) {PDCP layer \\ switches the RAT};
    \draw [-{Latex}](10, 7.2) -- node[sloped, anchor=center, above] {Send Switch to LTE} (5, 6.9);
    \draw [-{Latex}, dashed] (10,7.0) -- node[sloped, anchor=west, above] {Send RRC Connection Switch$\quad\quad\quad\quad\quad\quad\quad\quad\quad\quad\quad\quad\quad$} (0, 6.4);

    \draw [-{Latex}] (5, 6.5) -- node[sloped, anchor=center, below] {Forward RLC buffers} (10, 6.2);

    \filldraw[fill=midgrey!20] (-0.2,5.4) rectangle (3.4,4.4);
    \node [align=center] (ueSwitch) at (1.5,4.9) {PDCP layer \\ switches the RAT};

  \end{tikzpicture}
  \caption{Switch from mmWave RAT to LTE RAT.}
  \label{fig:mmToLte}
\end{subfigure}
\caption{Proposed RAT switch  procedures.}
\label{fig:switch}
\end{figure*}

Table \ref{tab:BF_arch} reports the delay $D$ for different configurations of a system with $N_{\rm UE} = 8$ and $N_{\rm eNB} = 16$ directions required to collect each instance of the CRT at the LTE eNB side by implementing the framework described above. For example, by implementing a hybrid BF with $L=2$ RF chains, the eNB can simultaneously receive through $L=2$ directions at the same time \cite{sun2014mimo} so the overall delay is $D=12.8$~ms. 

From the protocol stack point of view, unlike in~\cite{3GPP_DC}, both Radio Access Technologies (RATs) have a complete RRC layer  in the eNBs and in the UE. This allows a larger flexibility, since the design of the mmWave RRC layer can be decoupled from that of the LTE stack. 
Moreover, the LTE RRC is used for the management of the LTE connection but also to send and receive commands related to DC, while the mmWave RRC is used to manage only the mmWave link and the reporting of measurements to the coordinator. 
The choice of using a dedicated RRC link for the secondary eNB is motivated by the desire to reduce the latency of control commands, since it avoids the encoding and transmission of the control PDUs of the secondary cell to the master cell. The mmWave signaling radio bearers are used only when a connection to LTE is already established, and this can offer a ready backup in case the mmWave link suffers an outage. 

\subsection{User Plane (PDCP Layer Integration)}
In a DC architecture, the layer at which the LTE and the mmWave protocol stacks merge is called \textit{integration layer}. In this paper we propose the PDCP layer as the integration layer. In fact, it allows a non co-located\footnote{MmWave eNBs will be deployed more densely than already installed LTE eNBs, therefore it would be costly to have only co-located cells. Moreover a high density of LTE eNBs would decrease the effectiveness of the coverage layer. Finally, the PDCP layer can also be deployed in the core network, in a new node (\textit{coordinator}), which can be a gateway for a cluster of LTE eNBs and the mmWave eNBs under their coverage, or can be deployed in a macro LTE cell.} deployment, since synchronization among the lower layers is not required, 
and it does not impose any constraint on the design of mmWave PHY to Radio Link Control (RLC) layers, so that a clean slate approach can be used to address mmWave specific issues and reach 5G performance requirements. 

For each bearer, a PDCP layer instance is created in the LTE eNB and interfaced with the X2 link that connects to the remote eNB. Local and remote RLC layer instances are created in the LTE and mmWave eNB, respectively. The packets are routed from the S-GW to the LTE eNB, and once in the PDCP layer they are forwarded either to the local LTE stack or to the remote mmWave RLC. If there exists at least one mmWave eNB not in outage and the UE is connected to it, then the mmWave RAT is chosen, i.e., the LTE connection is used only when no mmWave eNB is available. This choice is motivated by the fact that the theoretical capacity of the mmWave link is greater than that of the LTE link~\cite{akoum2012coverage}, and that the LTE eNB will typically serve more users than the mmWave eNBs; however, when the mmWave eNBs are in outage (as it may happen in a mmWave context) and would therefore provide zero throughput to their users, an LTE link may be a valid fallback alternative to increase the robustness of the connection.
In addition, integration at the PDCP layer ensures ordered delivery of packets to the upper layers, which is useful in handover circumstances.

\subsection{Dual Connectivity-aided Network Procedures}
\label{sec:handover}

The DC framework allows to design network procedures that are faster than the standard standalone hard handover (HH), thus improving the mobility management in mmWave networks. The standalone HH architecture will be the baseline for the performance evaluation of Sec.~\ref{sec:results}: the UE is connected  to either the LTE  or the mmWave RAT and, in order to switch from one to the other, it has to perform a complete handover, or, if the mmWave connectivity is lost, an initial access to LTE from scratch. Besides, in order to perform a handover between mmWave eNBs, the UE has to interact with the MME in the core network, introducing additional delays. The DC architecture, instead, allows to perform fast switching between the LTE and  mmWave RATs and Secondary Cell Handover (SCH) across mmWave eNBs.

The fast switching procedure is used when all the mmWave eNBs for a certain UE are in outage. Since the handling of the state of the user plane for both the mmWave and the LTE RATs is carried out by the LTE RRC, it is possible to correctly modify the state of the PDCP layer and perform a switch from the mmWave  to the LTE RAT. The proposed switch procedure, shown in Fig.~\ref{fig:switch}, simply requires an RRC message (RRC Connection Switch command) to the UE, sent on the LTE link, and a notification to the mmWave eNB via X2 if the switch is from mmWave to LTE, in order to forward the content of the RLC buffers to the LTE eNB.

\begin{figure}[t]
\centering
  \begin{tikzpicture}[font=\sffamily\small, scale=0.5, every node/.style={scale=0.5}]
    \coordinate (UeBegin) at (0,10);
    \coordinate (UeEnd) at (0,0);
    \coordinate (mmWaveEnb1Begin) at (5,10);
    \coordinate (mmWaveEnb1End) at (5,0);
    \coordinate (mmWaveEnb2Begin) at (10,10);
    \coordinate (mmWaveEnb2End) at (10,0);
    \coordinate (lteEnbBegin) at (15,10);
    \coordinate (lteEnbEnd) at (15,0);
    
    \node [above of=UeBegin, yshift=-15pt] (UElabel) {UE};
    \node [above of=mmWaveEnb1Begin,yshift=-15pt] (mmWaveEnbLabel) {Source mmWave eNB $i$};
    \node [above of=mmWaveEnb2Begin,yshift=-15pt] (mmWaveEnbLabel) {Target mmWave eNB $j$};
    \node [above of=lteEnbBegin,yshift=-15pt] (lteEnbLabel) {LTE eNB};

    \begin{scope}[on background layer]
      \draw [thick, black] (UeBegin) -- (UeEnd);
      \draw [thick, black] (mmWaveEnb1Begin) -- (mmWaveEnb1End);
      \draw [thick, black] (mmWaveEnb2Begin) -- (mmWaveEnb2End);
      \draw [thick, black] (lteEnbBegin) -- (lteEnbEnd);
    \end{scope}

    \filldraw[fill=midgrey!20] (11.7,9.8) rectangle (15.2,8.8);
    \node [align=center] (triggerHO) at (13.5,9.3) {Coordinator triggers\\HO to $j$};

    \draw [-{Latex}] (15, 8.4) -- node[sloped, anchor=center, above] {Send Secondary cell Handover Request} (5, 7.8);
    \draw [-{Latex}] (5, 7.2) -- node[sloped, anchor=center, above] {Send Handover Request} (10, 6.9);
    \draw [-{Latex}] (10, 6.4) -- node[sloped, anchor=center, above] {Send Handover Request ACK} (5, 6.1);
    \draw [-{Latex}, dashed] (5, 5.8) -- node[sloped, anchor=center, above] {RRC Connection Reconf.} (0, 5.5);

    \draw [-{Latex}] (5, 5.7) -- node[sloped, anchor=center, above] {Forward RLC buffers} (10, 5.4);

    \filldraw[fill=midgrey!20] (-0.2, 5.1) rectangle (10.2,4.1);
    \node (initialLte) at (5,4.6) {LTE-aided Non Contention Based RA};

    \draw [-{Latex}, dashed] (0, 3.7) -- node[sloped, anchor=center, above] {RRC Connection Reconf. Completed} (10, 3.1);
    
    \draw [-{Latex}] (10, 2.7) -- node[sloped, anchor=center, above] {SCH completed} (15, 2.4);

    \filldraw[fill=midgrey!20] (11.7,2.0) rectangle (15.2,1.0);
    \node [align=center] (ps) at (13.5,1.5) {Path Switch from\\source to target};

    \draw [-{Latex}] (15, 0.6) -- node[sloped, anchor=center, above] {Remove UE Context} (10, 0.3);

  \end{tikzpicture}
  \caption{Secondary cell Handover procedure (SCH).}
  \label{fig:sho}
\end{figure}
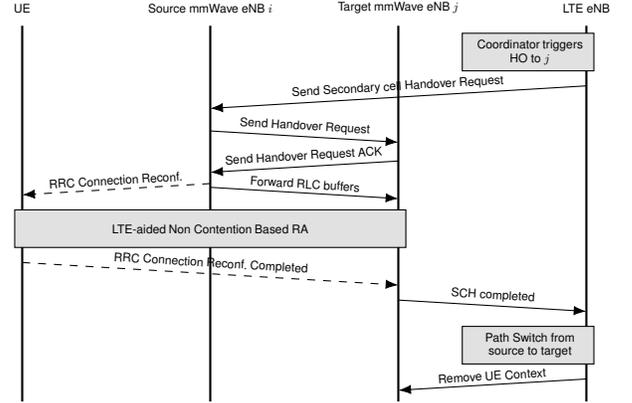

The DC solution therefore allows to have an uninterrupted connection to the LTE anchor point. However, it is possible to switch from a secondary mmWave eNB to a different mmWave eNB with a procedure which is faster than a standard intra RAT handover, since it does not involve the interaction with the core network. The Secondary Cell Handover procedure is shown in Fig.~\ref{fig:sho}. 
The Random Access (RA) procedure \cite{LTE_book} is aided by the measurement collection framework described in Section \ref{sec:measurement_framework}, which allows to identify the best beam to be used by the UE and avoids the need for the UE to perform an initial beam search.
Moreover, if the UE is capable of maintaining timing control with multiple mmWave eNBs, the RA procedure in the target mmWave eNB can be skipped.

We also propose an algorithm for SCH, based on the SINR measurements reported by the mmWave eNBs to the coordinator and on a threshold in time (TTT). 
When a mmWave eNB has a better SINR than the current one (and neither of the two is in outage), the LTE coordinator checks for TTT seconds if the condition still holds, and eventually  triggers the SCH. Notice that, if during the TTT the SINR of a third cell becomes better than that of the target cell by less than 3 dB, the handover remains scheduled for the original target eNB, while, if the original cell SINR becomes the highest, then the SCH is canceled. The TTT is computed in two different ways. With the \textit{fixed} TTT option it always has the same value\footnote{This approach recalls the standard HO for LTE networks.} (i.e., $f_{TTT} = 150$ ms), while for the \textit{dynamic} TTT case we introduce a dependency on the difference $\Delta$ between the SINRs of the best and of the current cell:
\beq
  f_{TTT}(\Delta) = TTT_{max} - \frac{\Delta - \Delta_{min}}{\Delta_{max} - \Delta_{min}}(TTT_{max} - TTT_{min})
\eeq
so that the actual TTT value is smaller when the difference in SINR between the current eNB and the target is higher.  
The parameters that were used in the performance evaluation carried out in this paper are $TTT_{max} = 150$~ms, $TTT_{min} = 25$~ms, $\Delta_{min}=3$~dB, $\Delta_{max}=8$~dB.

Finally, if at a given time all the mmWave eNBs are in outage, then the UE is instructed to switch to the LTE eNB. If instead only the current mmWave eNB is in outage,  the UE immediately performs a handover to the best available mmWave eNB, without waiting for a~TTT. 

\section{Performance Evaluation Framework}

In order to assess the performance of the proposed DC architecture with respect to the traditional standalone hard handover (HH) baseline 
we use ns--3-based system level simulations, based on the DC framework described in~\cite{simutoolspolese}. This approach has the advantage of including many more details than 
would be allowed by an analytical model (which, for such a complex system, would have to introduce many simplifying assumptions), and makes it possible to evaluate the system performance accounting for realistic (measurement-based) channel behaviors and detailed (standard-like) protocol stack implementations. 
The higher layers of the LTE and mmWave protocol stacks are an extension of their respective counterparts of the ns--3 LTE module~\cite{lena}. The PHY and MAC layers of the mmWave stack are instead the ones described in~\cite{mmWaveSim,phy5g}. The LTE classes and the mmWave PHY and MAC layers were extensively modified in order to support the dual connectivity framework described in Section~\ref{sec:MCP}.

The mmWave physical layer is based on a Time Division Duplexing (TDD) frame structure~\cite{duttaTDD}, which can be configured on many parameters. 
The MAC layer offers (i) scheduling according to a TDMA scheme with  variable slot duration, which makes it possible to increase the efficiency of the resource utilization and to accomodate transport blocks of different sizes (TDMA on top of a TDD scheme is one of the options for 5G MAC and PHY layer design~\cite{duttaTDD}), (ii) adaptive modulation and coding, and (iii) Hybrid ARQ retransmissions. 

The source code of the DC framework is publicly available\footnote{\url{https://github.com/nyuwireless/ns3-mmwave/tree/new-handover}}, as well as the ns--3 script (\texttt{mc-example-udp.cc}) used for the simulation scenario considered in this paper.

\begin{figure*}[t!]
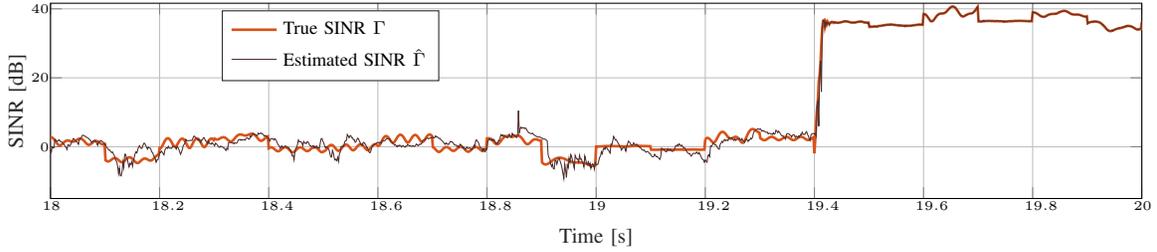

\centering
\iftikz
  \setlength\fwidth{0.8\textwidth}
  \setlength\fheight{0.15\textwidth}
  \input{./figures/example_SINR.tex}
\else
 \includegraphics[trim= 0cm 0cm 0cm 0cm , clip=true, width= 0.9\textwidth]{figures/example_SINR}
 \fi
 \caption{SINR evolution, with respect to a specific mmWave eNB in the network, whose samples are collected every $D=1.6$~ms, according to the measurement framework described in Section \ref{sec:measurement_framework}. Each sample is obtained by following the semi-statistical channel model proposed in \cite{EW2016} and explained in this section. The red line is referred to the true SINR trace $\Gamma$, while the black line is referred to its estimate $\bar{\Gamma}$, after noise and a first-order filter are applied to the true SINR $\Gamma$.}
 \label{fig:example_SINR}
\end{figure*}

\subsection{Semi-Statistical Channel Model}
The channel model is based on recent real-world measurements at $28$~GHz in New York City, to provide a realistic assessment of mmWave micro and picocellular networks in a dense urban deployment \cite{Mustafa,Samimi2015Prob,MacCartney2015Wideband,samimi2015chanmod}.
Unfortunately, most of the studies have been performed in stationary locations with minimal local blockage, making it difficult to estimate the rapid channel dynamics that affect a realistic mmWave scenario.
Dynamic models such as \cite{eliasi2015stochastic} do not yet account for the
spatial characteristics of the channel.

Measuring a wideband spatial channel model with dynamics is not possible with our current experimental equipment, as such measurements would require that the transmitting and receiving directions be swept rapidly during the local blocking event. Since our available platform relies on horn antennas mounted on mechanically rotating gimbals, such rapid sweeping is not possible.

In this work, we follow the alternate approximate \textit{semi-statistical} method proposed in \cite{EW2016} to generate realistic dynamic models for link evaluation:

\begin{itemize}
\item[(i)] We first randomly generate the statistical parameters of the mmWave channel, according to \cite{Mustafa} and \cite{samimi2015chanmod}, which would reflect the characteristics of a stationary ground-level mobile with no local obstacles.
\item[(ii)] Since there are no statistical models for the blocking dynamics, local blocking events are measured experimentally and modulated on top of the static parameters, in case an obstacle is physically deployed through the path that links the UE to one of the mmWave eNBs\footnote{ An important simplification  is that we assume that the local blockage  equally attenuates all paths, which may not always be realistic. For example, a hand may block only paths in a limited number of directions. However, in any fixed direction, most of the power is contributed only by paths within a relatively narrow beamwidth and thus the approximation that the paths are attenuated together may be reasonable.}.
\end{itemize}

Handover decisions described in Section \ref{sec:handover} are based on the SINR values saved in the CRT, built at the coordinator's side. Specifically, the SINR between a mmWave eNB$_j$ and a test UE can be computed in the following way:
\begin{equation}
\text{SINR}_{ j,\rm UE} = \frac{\frac{P_{\rm TX}}{PL_{ j,\rm UE}}G_{ j,\rm UE}}{\sum_{k\neq j}\frac{P_{\rm TX}}{PL_{ k,\rm UE}}G_{ k,\rm UE}+W_{\rm tot}\times N_0}
\vspace{0.5cm}
\label{eq:SINR}
\end{equation}
where $G_{ i,\rm UE}$ and $PL_{ i,\rm UE}$ are the beamforming  gain and the pathloss  obtained between eNB$_{ i}$ and the UE, respectively, $P_{\rm TX}$ is the transmit power and $ W_{\rm tot}\times N_0$ is the thermal noise power.

In the following, we describe in detail how the real experiments in common blockage scenarios are combined with the outdoor statistical model for ns--3, to get a realistic expression for the SINR samples which takes into account the dynamics experienced in a mmWave channel.\\

\subsubsection{\textbf{MmWave Statistical Channel Model}}

The parameters of the mmWave channel that are used to generate the  time-varying channel matrix $\mathbf{H}$ include: (i)  spatial clusters, described by central azimuth and elevation angles; (ii) fractions
of power;
(iii) angular beamspreads; and 
(iv) small-scale fading,
which models every small movement (e.g., a slight variation of the handset orientation) and is massively affected by the Doppler shift and the real-time position (AoA, AoD) of the UE, which may change very rapidly, especially in dense and high-mobility scenarios (for this reason, we chose to adapt the channel's small scale fading parameters as frequently as possible, that is once every time slot of 125 $\mu s$).

These parameters are defined and explained in \cite{samimi2015chanmod,Mustafa}, while a complete description of the channel model can be found in \cite{MCJournal2016}.
Notice that, following the approach of~\cite{mmWaveSim}, the large scale fading parameters of the $\mathbf{H}$ matrix are updated every 100~ms, to simulate a sudden change of the  link quality. 

The pathloss is defined as
$PL(d)[dB] = \alpha + \beta 10 \log_{10}(d)$,
where $d$ is the distance between the receiver and the transmitter and the values of the parameters $\alpha$ and $\beta$ are given in \cite{Mustafa}.
In case an obstacle is obstructing the path that links the UE and a specific mmWave eNB in the network, a Non-Line-Of-Sight (NLOS) pathloss state is emulated by superimposing the experimentally measured blockage traces to the statistical realization of the channel, as explained in Section \ref{sec:finalSINRtrace}.

When just relying on the statistical characterization of the mmWave channel, the SINR expression obtained by applying Equation \eqref{eq:SINR} assumes a baseline Line-Of-Sight (LOS) pathloss where no local obstacles affect the propagation of the signal. 
In the next paragraph, a channel sounding system is presented for measuring the dynamics of the blockage.\\

\subsubsection{\textbf{Measurement of Local Blockage}}
The key challenge in measuring the dynamics of local blockage is that we need relatively fast measurements. To perform these fast measurements, we used a high-bandwidth baseband processor, built on a PXI (a rugged PC-based platform for measurement and automation systems) from National Instruments, which engineers a real-world mmWave link. A detailed description of the experimental testbed can be found in \cite{EW2016} and \cite{aditya}.

Using this system, the experiments were then conducted by placing moving obstacles (e.g., a person walking or running) between the transmitter and the receiver, and continuously collecting Power Delay Profile (PDP) samples during each blocking event\footnote{PDPs were measured at a rate of one PDP every 32~$\mu$s but, since we found that the dynamics of the channel varied considerably slower than this rate, we decimated the results by a factor of almost four, recording one PDP every about 125~$\mu$s, that matches the slot duration of the ns--3 framework.}.  
The experiments show that obstacles can cause up to 35-40~dB of attenuation with respect to the LOS baseline SINR values, and this local blocking attenuation factor is thus used to modulate the time-varying channel response from the statistical channel model.\\

\subsubsection{\textbf{Final Semi-Statistical SINR Trace}}
\label{sec:finalSINRtrace}

Once a statistical instance of SINR$_{j,\rm{UE}}$ is obtained from Equation \eqref{eq:SINR}, a raw estimate of the real SINR at the UE is derived by superimposing the local blocking dynamics measured experimentally, when an obstacle is physically present in the path between that UE and eNB$_{j}.$ In particular, we denote by $\Gamma_{\mathrm{stat},j}$ the maximum static SINR between the eNB$_j$ and the UE receiver, when assuming that no local obstacles are present. Then the maximum wideband SINR  when also considering a dynamic model for the link evaluation (that is the value inserted in the $j$-th column of the CRT, at a specific time instant) is obtained as:
\begin{equation}
\Gamma_j =\begin{cases}  \Gamma_{\mathrm{stat},j} & \parbox[t]{.65\columnwidth}{if no obstacles are in the path between  UE and  eNB$_j$ \textit{(LOS  condition)}} \\  \delta + \Gamma_{\mathrm{stat},j}& \parbox[t]{.65\columnwidth}{if an obstacle is  in the path between  UE and eNB$_j$ \textit{(NLOS  condition)}} \end{cases}
\end{equation}
where $ \delta$ is a scaling factor that accounts for the SINR drop measured experimentally in various blocking scenarios and collected using the instrumentation described in the previous paragraph.

This final semi-statistical SINR trace is composed of samples of $\Gamma_j$ generated every $125 \: \, \mu s$ (from both the statistical trace and the experimental measurements). 
Finally, according to Section \ref{sec:MCP}, the HO decisions are made once the coordinator has built a CRT, that is every $D$ seconds.
Thus, the original SINR trace has been downsampled, keeping just one sample every $D$~seconds.

\subsection{SINR Filtering}
The mmWave eNBs  estimate the wideband SINR $\Gamma_j$ from the sounding reference signals that are transmitted by the UE and are collected by each mmWave eNB, to build a CRT at the coordinator's side\footnote{The estimation of the channel is relatively straightforward in 3GPP LTE \cite{LTE_book,schwarz2010calculation} and is based on the cell reference signal (CRS) that is continuously and omnidirectionally sent from each eNB. However, a CRS will likely not be available in mmWave systems, since downlink transmissions at mmWave frequencies will be directional and specific to the UE \cite{EW2016}.}.
However, the \textit{raw estimate} of the SINR $\hat{\Gamma}_j$, that is what is really measured in a realistic communication system, may deviate from  $\Gamma_j$ due to noise (whose effect can be very significant when considering low SINR regimes). 
To reduce the noise, $\hat{\Gamma}_j$ is filtered, producing a time-averaged SINR trace $\bar{\Gamma}_j$. According to \cite{EW2016}, a simple first-order filter can properly restore the desired SINR stream and perform reliable channel estimation even without designing  more complex and expensive adaptive nonlinear filters. Therefore, $\bar{\Gamma}_j$ is obtained as
\beq \label{eq:gamFO}
    \bar{\Gamma}_i = (1-\eta) \bar{\Gamma}_{i-1} + \eta \hat{\Gamma}_i,
\eeq
for some constant $\eta \in (0,1)$  chosen in order to minimize the estimation error $e_i = |\Gamma_i - \bar{\Gamma}_i|^2$.

As an example, the SINR trace in Fig.  \ref{fig:example_SINR} (whose samples are collected every $D=1.6$~ms) is obtained by following the semi-statistical channel model proposed in \cite{EW2016} and explained in this section.
For time $t<19.4$, the UE is in a NLOS pathloss condition with respect to its eNB, therefore a scaling factor $\delta$ measured experimentally is applied to the statistical trace to account for the dynamics of the local blockage. For time $t>19.4$, the UE enters a LOS state until the end of the simulation. 
The SINR collapses and spikes within the trace (i.e., at times $t=18.1$ or $t=18.9$)  are mainly caused by the update of the large scale fading parameters of the statistical mmWave channel, while the rapid fluctuations of the SINR are due to the adaptation of the small scale fading parameters of \textbf{H} (and mainly to the Doppler effect experienced by the moving user).
Finally, the red and black lines are referred to the true measured SINR trace $\Gamma$ and its estimate $\bar{\Gamma}$ (after the noise and a first-order filter are applied), respectively. We observe that, for low SINR regimes, $\bar{\Gamma}$ presents a noisy trend but appears still similar to the original trace while, when considering good SINR regimes (e.g., when the UE is in LOS), the estimated trace almost overlaps with  its measured original version.

\subsection{Simulation Parameters}
\label{sec:simparams}

\begin{figure}[t!]
\centering
\setlength{\belowcaptionskip}{-0.3cm}

\begin{tikzpicture}[font=\sffamily, scale=0.45, every node/.style={scale=0.45}]
  \centering

    \node[anchor=south west,inner sep=0] (image) at (0,0) {\includegraphics[width=0.9\textwidth]{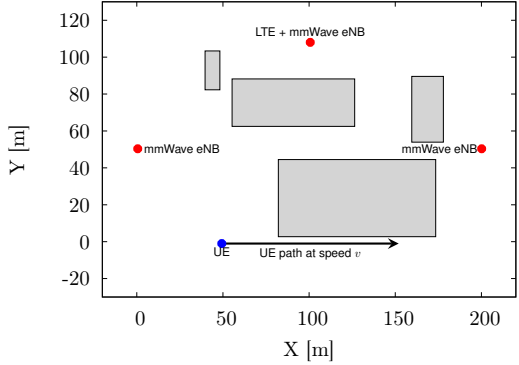}};
    \begin{scope}[x={(image.south east)},y={(image.north west)}]
        \filldraw[red,ultra thick] (0.248,0.555) circle (2pt);
        \node[anchor=west] at (0.252,0.555) (mm1label) {mmWave eNB};
        
        \filldraw[red,ultra thick] (0.87,0.555) circle (2pt);
        \node[anchor=east] at (0.87,0.555) (mm2label) {mmWave eNB};
        \filldraw[red,ultra thick] (0.56,0.83) circle (2pt);
        \node[anchor=south] at (0.56,0.835) (ltelabel) {LTE + mmWave eNB};

        \draw[sarrow] (0.4, 0.31) -- node[anchor=north] {UE path at speed $v$} (0.72, 0.31);

        \filldraw[blue,ultra thick] (0.4, 0.31) circle (2pt);
        \node[anchor=north] at (0.4, 0.31) (ltelabel) {UE};
    \end{scope}   
\end{tikzpicture}
\caption{Random realization of the simulation scenario. The grey rectangles are $4$ randomly deployed non-overlapping buildings.}
\label{fig:scenario}
\end{figure}

The reference scenario (for which one example of random realization is presented in Fig.  \ref{fig:scenario}) is a typical urban grid having area $200 \times 115$ meters, where $4$ non-overlapping buildings of random size and height are deployed, in order to randomize the channel dynamics (in terms on LOS-NLOS transitions) for the moving user. Three mmWave eNBs are located at coordinates {eNB$_{2}=(0;50)$}, {eNB$_{3}=(200;50)$} and {eNB$_{4}=(100;110)$}, at a height of $10$ meters. The LTE {eNB$_{1}$} is co-located with eNB$_4$.
We consider a single UE that is at coordinates {$(50;-5)$} at the beginning of the simulation. 
It then moves along the x-axis at speed $v$ m/s, until it arrives in position {$(150;-5)$}. 
The simulation duration $T_{\rm sim}$ therefore depends on the UE speed $v$ and is given by $T_{\rm sim} = \frac{l_{\rm path}}{v}=20$~s, where $l_{\rm path}=100$~m is the length of the path of the UE during the simulation and the default value of the mobile speed has been taken to be $v=5$ m/s. 
 
\begin{table}[!t]
\setlength{\belowcaptionskip}{-0.3cm}
\renewcommand{\arraystretch}{1}
\footnotesize
\centering
\begin{tabular}{@{}lll@{}}
\toprule
Parameter & Value & Description \\ \midrule

 mmWave $W_{\rm tot}$ & $1$ GHz & Bandwidth of mmWave eNBs\\

mmWave $f_{\rm c}$ & $28$ GHz & mmWave carrier frequency \\

mmWave $P_{\rm TX}$ & $30$ dBm & mmWave transmission power \\

LTE $W_{\rm tot}$ & $20$ MHz & Bandwidth of the LTE eNB\\

LTE $f_{\rm c}$ & $2.1$ GHz & LTE  carrier frequency \\

LTE DL  $P_{\rm TX}$ & $30$ dBm & LTE DL transmission power \\

LTE UL  $P_{\rm TX}$ & $25$ dBm & LTE UL transmission power \\

 NF  & $5$ dB & Noise figure \\

$\Gamma_{out}$ & $ -5$ dB &  Minimum SINR threshold \\

eNB antenna & $8 \times 8$  & eNB UPA MIMO array size  \\

UE antenna & $4 \times 4$ & UE UPA MIMO array size\\

$N_{\rm eNB}$& $16$  & eNB scanning directions  \\

$N_{\rm UE}$& $8$  & UE scanning directions  \\

$T_{\rm sig}$ & $10 \, \: \mu s$& SRS duration \\

$\phi_{\rm ov}$ & $5\%$ & Overhead\\

$T_{\rm per}$ & $200 \: \mu$s & Period between SRSs \\

$v$ & $5$ m/s & UE speed\\

$B_{\rm RLC}$ & 10 MB & RLC buffer size\\

$D_{\rm X2}$ & $1$ ms & One-way delay on X2 links\\

$D_{\rm MME}$ & $10$ ms & One-way MME delay \\ 

$T_{\rm UDP}$ & $\{20,80\} \:\mu s$ & UDP packet interarrival time \\

$s_{\rm UDP} $ & 1024 byte & UDP payload size\\

$D$ & $\{1.6,12.8,25.6\}$ ms  & CRT intergeneration delay\\
\bottomrule
\end{tabular}
\caption{Simulation parameters.}
\label{tab:params}
\end{table}

\begin{figure*}[t]
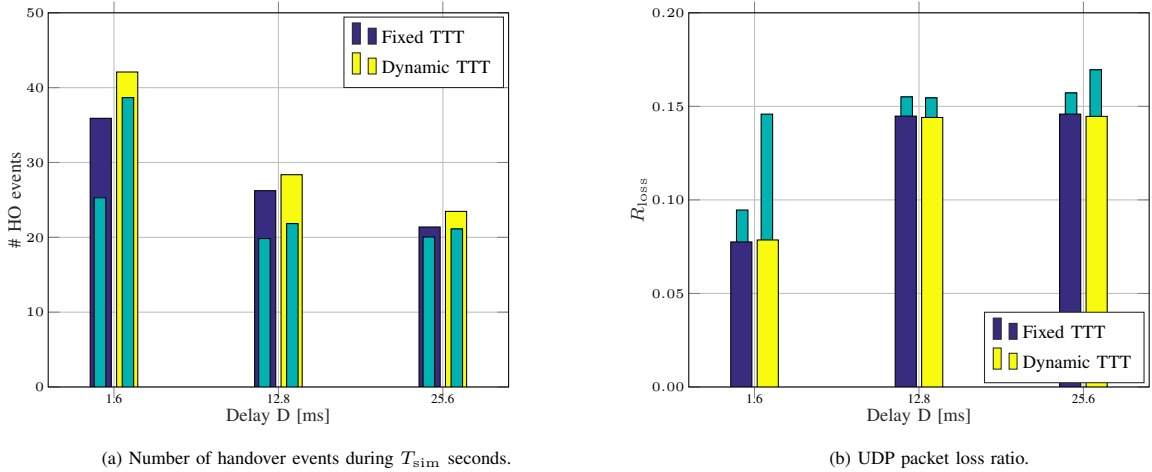

  \centering
  \begin{subfigure}[t]{0.45\textwidth}
    \setlength{\belowcaptionskip}{0cm}
    \iftikz
    \setlength\fwidth{0.75\textwidth}
    \setlength\fheight{0.61\textwidth}
    \input{./figures/handovers_10MB.tex}
    \else
    \includegraphics[width = \textwidth]{handovers_10MB}
    \fi
    \caption{Number of handover events during $T_{\rm sim}$ seconds.}
  \end{subfigure}
  \begin{subfigure}[t]{0.45\textwidth}
    \setlength{\belowcaptionskip}{0cm}
    \iftikz
    \setlength\fwidth{0.75\textwidth}
    \setlength\fheight{0.61\textwidth}
    \input{./figures/loss_10MB.tex}
    \else
    \includegraphics[width = \textwidth]{loss_10MB}
    \fi
      \caption{UDP packet loss ratio.}
  \end{subfigure}
  \caption{Average number of handover events and packet loss ratio, for different values of the delay $D$, for a fixed and dynamic TTT HO algorithm. Narrow bars refer to a hard handover configuration, while wide colored bars refer to a dual connectivity implementation. The RLC buffer size is $B=10$ MB and the interarrival packet time is $T_{\rm UDP}=20 \: \mu s$.}
  \label{fig:n_HO} 
\end{figure*}
 
Our results are derived through a Monte Carlo approach, where multiple independent simulations are repeated, to get different statistical quantities of interest. In each experiment: (i) we randomly deploy the obstacles; (ii) we apply the measurement framework described in Section \ref{sec:measurement_framework} to collect one CRT every $D$ seconds; and (iii) we eventually employ one of the  HO algorithms presented in Section \ref{sec:handover}.

The goal of these simulations is to assess the difference in performance between a system using dual connectivity, with fast switching and SCH, and another where hard handover (HH) is used, for different values of $D$, i.e., when varying the periodicity of the CRT generation at the LTE eNB side.
Indeed the comparison between these two configurations can be affected by several parameters, which are based on realistic system design considerations and are summarized in Table \ref{tab:params} \cite{simutoolspolese}. On the other hand, the performance of the two options does not depend on the interference, since its impact is similar in both schemes.
The value of the delay to the MME node ($D_{\rm MME}$) is chosen in order to model both the propagation delay to a node which is usually centralized and far from the access network, and the processing delays of the MME server.
We also model the additional latency $D_{\rm X2}$ introduced by the X2 connections between each pair of eNBs, which has an impact on (i) the forwarding of PDCP PDUs from the LTE eNB to the mmWave ones; (ii) the exchange of control messages for the measurement reporting framework and (iii) the network procedures which require coordination among eNBs. Thus, the latency $D_{\rm X2}$ may delay the detection at the LTE eNB coordinator of an outage with respect to the current mmWave link. In order to avoid performance degradation, the value of $D_{\rm X2}$ should be smaller than 2.5 ms, as recommended by~\cite{x2lat}.

We consider an SINR threshold $\Gamma_{out} = -5$ dB, assuming that, if $\bar{\Gamma}_j(t) < \Gamma_{out}$, no control signals are collected by eNB$_j$ at time $t$ when the UE is transmitting its SRSs. Reducing  $\Gamma_{out}$ allows the user to be potentially found by more suitable mmWave cells, at the cost of designing more complex (and expensive) receiving schemes, able to detect the intended signal in more noisy channels. 
eNBs are equipped with a Uniform Planar Array (UPA) of $8 \times 8$ elements, which allow them to steer beams in $N_{\rm eNB}=16$ directions, whereas UEs have a UPA of $4 \times 4$ antennas, steering beams through $N_{\rm UE}=8$ angular directions. 

The behavior of the UDP transport protocol (whose interarrival packet generation time is~$T_{\rm UDP}$) is tested, to check whether our proposed dual connectivity framework  offers good resilience in mobility scenarios. Only downlink traffic is considered.

\section{Results And Discussion}
\label{sec:results}

In this section, we present some results that have been derived for the scenario presented in Section \ref{sec:simparams}. Different configurations have been compared in terms of packet loss, latency, PDCP throughput, RRC and X2 traffic in order to:
\begin{itemize}
\item[i)] compare DC with fast switching and SCH versus the traditional standalone hard handover architectures;
\item[ii)] compare the performance of the dynamic and the fixed TTT HO algorithms;
\item[iii)] validate our proposed measurement reporting system varying the CRT intergeneration periodicity $D$ and the UDP interarrival packet time $T_{\rm UDP}$. 
\end{itemize}

\begin{figure*}[t]
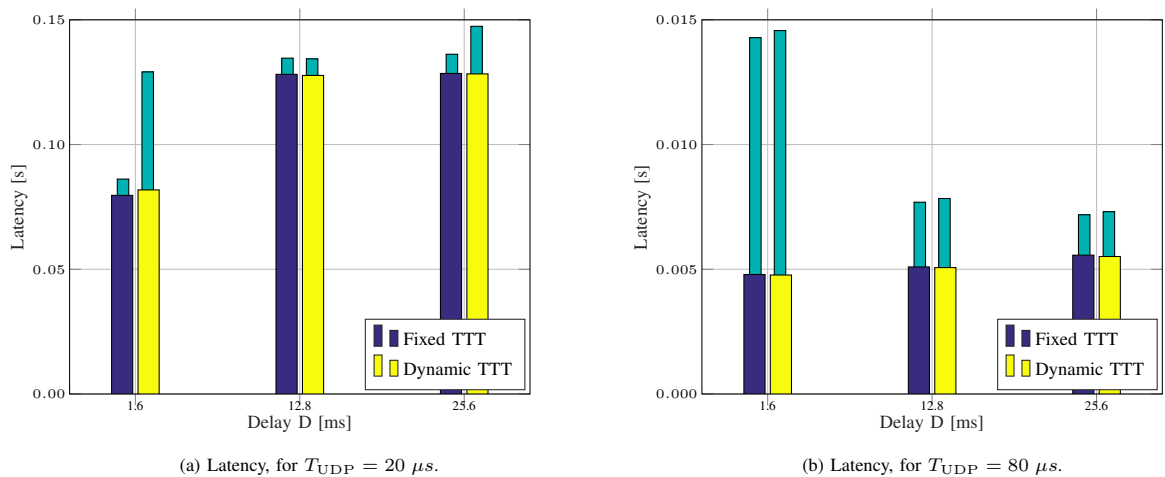

  \centering
  \begin{subfigure}[t]{0.45\textwidth}
    \iftikz
    \setlength{\belowcaptionskip}{0cm}
    \setlength\fwidth{0.75\textwidth}
    \setlength\fheight{0.61\textwidth}
    \input{./figures/latency_10MB.tex}
    \else
    \includegraphics[width = \textwidth]{latency_10MB}
    \fi
    \caption{Latency, for $T_{\rm UDP} = 20 \; \mu s$.}
  \end{subfigure}
  \begin{subfigure}[t]{0.45\textwidth}  
    \setlength{\belowcaptionskip}{0cm}
    \iftikz
    \setlength{\belowcaptionskip}{0cm}
    \setlength\fwidth{0.75\textwidth}
    \setlength\fheight{0.61\textwidth}
    \input{./figures/latency_10MB_80us.tex}
    \else  
    \includegraphics[width = \textwidth]{latency_10MB_80us}
    \fi
    \caption{Latency, for $T_{\rm UDP} = 80 \; \mu s$.}
  \end{subfigure}
  \caption{Average latency, for different values of the delay $D$ and the UDP packet interarrival time $T_{\rm UDP}$, for a fixed and dynamic TTT HO algorithm. Narrow bars refer to a hard handover configuration, while wide colored bars refer to a dual connectivity implementation. The RLC buffer size is $B_{\rm RLC}=10$ MB.}
  \label{fig:latency} 
\end{figure*}

\subsection{Packet Loss and Handover}
\label{sec:nHO}
In Fig. \ref{fig:n_HO}(a) we plot the \textit{average number of handover} (or switch) events. 
As expected, we notice that this number is much higher when considering the DC configuration. The reason is that, since the DC-aided fast switching and SCH procedures are faster than the traditional standalone hard handover, the UE has more chances to change its current cell and adapt to the channel dynamics in a more responsive way.
Moreover, when increasing the delay $D$, i.e., when reducing the CRT generation periodicity, the number of handovers reduces, since the UE may have fewer opportunities to update its serving cell, for the same simulation duration.
Finally, we see that a dynamic HO procedure requires, on average, a larger number of handover events, to account for the situations in which TTT$<150$ ms, when the UE may change its serving cell earlier than it would have done if a fixed TTT algorithm had been applied.

Another element to consider in this performance analysis is the \textit{packet loss ratio} $R_{\rm loss}$, plotted in Fig.  \ref{fig:n_HO}(b)\footnote{The presented figure has been obtained when setting $T_{\rm UDP}=20 \: \mu s$. We have also tested the configuration $T_{\rm UDP}=80 \: \mu s$, but we saw that, across the different realizations of the simulation, $R_{\rm loss}$ was zero, due to the fact that the UDP traffic injected in the system was sufficiently well handled by the buffer, with no overflow.}, and defined as the ratio between lost and sent packets, averaged over the $N$ different iterations for each set of parameters. Since the UDP source constantly injects packets into the system,
with interarrival time $T_{\rm UDP}$, it can be computed as 
$R_{\rm loss} = 1-r{T_{\rm UDP}}/{T_{\rm sim}}$
where $r$ is the total number of received packets and $T_{\rm sim}$ is the duration of each simulation.
We first notice that, with the use of the DC solution, fewer packets are lost. In fact, there are mainly two elements that contribute to the losses: 
(i) some UDP packets, which are segmented in the RLC retransmission buffer, cannot be reassembled at the PDCP layer and are therefore lost;
(ii) during handover, the target eNB RLC transmission buffer receives both the packets sent by the UDP application with interpacket interval $T_{\rm UDP}$ and the packets that were in the source eNB RLC buffer. If the latter is full, then the target eNB buffer may overflow and discard packets.

Both these phenomena are stressed by the fact that the standalone HH procedure takes more time than both the DC-aided fast switching and SCH procedures. 
Moreover, during a complete outage event, with the HH solution, until the UE has completed the Non Contention Based Random Access procedure with the LTE eNB, packets cannot be sent to the UE and must be buffered at the RLC layer. This worsens the overflow behavior of the RLC buffer. Instead, with fast switching, the UE does not need to perform random access, since it is already connected and, as soon as packets get to the buffer of the LTE eNB, they are immediately transmitted to the UE.

Fig. \ref{fig:n_HO}(b) also shows that the packet loss ratio increases when $D$ increases since, if handover or switch events are triggered less frequently, the RLC buffer occupancy increases, and so does the probability of overflow.

Finally, almost no differences are registered when considering a dynamic or fixed TTT HO algorithm, nor when increasing the CRT delay from $D=12.8$ ms to $D=25.6$ ms (this aspect will be explained in more detail later).

\subsection{Latency}
\label{sec:latency}

The latency is measured for each packet, from the time it leaves the PDCP layer of the LTE eNB to when it is successfully received at the PDCP layer of the UE. Therefore, it is the latency of only the correctly received packets, and it accounts also for the forwarding latency $D_{\rm X2}$ on the X2 link. Moreover, this metric captures the queuing time in the RLC buffers, and the additional latency that occurs when a switch or handover happens, before the packet is forwarded to the target eNB or RAT.  

Fig.~\ref{fig:latency} shows that the DC framework outperforms the standalone hard handover: in fact, as we pointed out in Section \ref{sec:nHO}, handovers (which dominate the HH configuration) take more time than the fast switching and SCH procedures, and therefore with DC the UE experiences a reduced latency and no service interruptions.
This result is even more remarkable when realizing that, from Fig. \ref{fig:n_HO}, the absolute number of handover (or switch) events is  higher when using DC: despite this consideration, the overall latency is still higher for a system where hard handover is implemented\footnote{The latency gap is even more remarkable when considering a dynamic TTT HO algorithm. In fact, although the UE experiences, on average, almost $15\%$ more handovers than in the fixed TTT configuration, the overall latency of the two configurations shown in Fig.~\ref{fig:latency} is comparable, due to the fact that with dynamic TTT some SCHs are more timely.}.

Furthermore, the latency increases as $D$ increases. In fact, when reducing the intergeneration time of the CRT, the UE is attached to a suboptimal mmWave eNB (or to the LTE eNB) for a longer period of time: this increases the buffer occupancy, thus requiring a stronger effort (and longer time) for forwarding many more packets to the new candidate cell, once the handover (or switch) is  triggered.
Finally, there are no remarkable differences between $D=12.8$ and $D=25.6$ ms. 

According to Fig. \ref{fig:latency}(b), the latency gap between the HH and DC configurations is much more impressive when considering $T_{\rm UDP}=80\:\mu s$. In fact, with this setup, the RLC buffer is empty most of the time and, when a handover (or a switch) is triggered, very few UDP packets need to be forwarded to the destination mmWave or LTE eNB, thus limiting the impact of~latency. 

We finally recall that, as already introduced in Section \ref{sec:handover}, the \emph{handover interruption time} (HIT, i.e., the time in which the user's connectivity is interrupted during the handover operations)  takes different values, according to the implemented handover scheme (either DC or HH).
When considering a switch to LTE, the HIT is negligible if a DC approach is used, since the UE is already connected to both the LTE and the mmWave RATs. There may be an additional forwarding latency for the switch from mmWave to LTE, which however is already accounted for in Fig.~\ref{fig:latency}. On the other hand, when referring to the baseline HH architecture, the UE has to perform a complete handover to switch from one RAT to the other, thus introducing a significant additional delay.
When considering the handover between mmWave eNBs, instead, the HIT is comparable for both the DC and the HH schemes. However, in the first case, the procedure does not involve any interaction with the core network and the UE is informed about the new mmWave eNB to handover to and the best angular direction to set through an LTE message (while, when choosing the HH configuration, the handover completion is postponed since the UE has to exhaustively scan again the angular space and perform a complete initial beam search to receive a connection-feedback message from the new serving mmWave eNB).
In general, the DC approach is thus preferred in terms of reduced interruption time too.

\begin{figure*}[t]
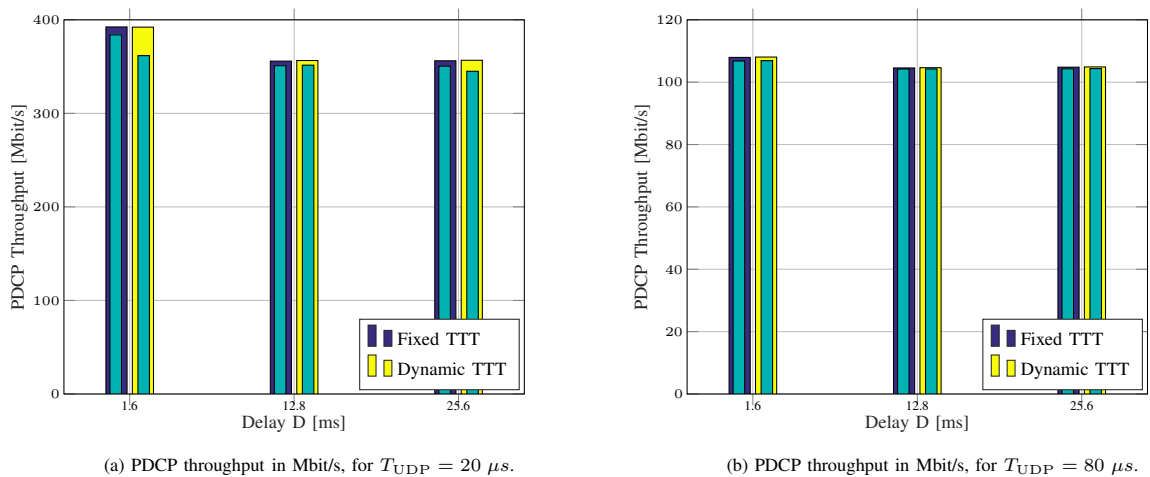

  \centering
  \begin{subfigure}[t]{0.45\textwidth}
    \setlength{\belowcaptionskip}{0cm}
    \iftikz
    \setlength{\belowcaptionskip}{0cm}
    \setlength\fwidth{0.75\textwidth}
    \setlength\fheight{0.61\textwidth}
    \input{./figures/th_10MB.tex}
    \else
    \includegraphics[width = \textwidth]{th_10MB}
    \fi
    \caption{PDCP throughput in Mbit/s,  for $T_{\rm UDP} = 20 \; \mu s$.}
  \end{subfigure}
  \begin{subfigure}[t]{0.45\textwidth}
    \setlength{\belowcaptionskip}{0cm}
    \iftikz
    \setlength{\belowcaptionskip}{0cm}
    \setlength\fwidth{0.75\textwidth}
    \setlength\fheight{0.61\textwidth}
    \input{./figures/th_10MB_80us.tex}
    \else
    \includegraphics[width = \textwidth]{th_10MB_80us}
    \fi
    \caption{PDCP throughput in Mbit/s, for $T_{\rm UDP} = 80 \; \mu s$.}
  \end{subfigure}
  \caption{Average PDCP throughput in Mbit/s, for different values of the delay $D$ and the UDP packet interarrival time $T_{\rm UDP}$, for fixed and dynamic TTT HO algorithm. Narrow bars refer to a hard handover configuration, while wide colored bars refer to a dual connectivity implementation. The RLC buffer size is $B_{\rm RLC}=10$ MB.}
  \label{fig:PDCPthroughput} 
\end{figure*}

\subsection{PDCP Throughput}

The throughput over time at the PDCP layer is measured by sampling the logs of received PDCP PDUs every $T_s = 5$ ms and summing the received packet sizes to obtain the total number of bytes received $B(t)$. Then the throughput $S(t)$ is computed in bit/s as $S(t) = B(t) \times 8/T_s$.
In order to get the mean throughput $S_{\rm PDCP}$ for a simulation, these samples are averaged over the total simulation time $T_{\rm sim}$, and finally over the $N$ simulations, to obtain the parameter $\mathbb{E}[S_{\rm PDCP}]$.
Notice that the PDCP throughput (which is mainly a measure of the rate that the radio access network can offer, given a certain application rate), is mostly made up of the transmission of new incoming packets, but it may also account for the retransmissions of already transmitted~ones.

In Fig.  \ref{fig:PDCPthroughput}, it can be observed that the throughput achievable with the dual connectivity solution is slightly higher than with hard handover. 
The reason is that, when relying on the LTE eNB for dealing with outage events, the UE experiences a non-zero throughput, in contrast to the hard handover configuration which cannot properly react to a situation where no mmWave eNBs are within reach.
Moreover, the difference in throughput increases as the application rate increases, in accordance with the results on packet loss described
in the previous section.

As expected, the PDCP throughput decreases as $D$ increases, since the CRT are generated less frequently and the beam pair between the UE and its serving mmWave eNB is monitored less intensively. This means that, when the channel conditions change (e.g., due to the user motion, to a pathloss condition modification or to the small and large scale fading parameters update), the communication quality is not immediately recovered and the throughput is affected by portions of time where suboptimal network settings are chosen.

Moreover, as pointed out in Section \ref{sec:latency}, we cannot see notable differences between the fixed and dynamic TTT HO procedures and between the $D=12.8$ and the $D=25.6$ ms CRT delays.
Also a lower UDP rate, according to Fig.~\ref{fig:PDCPthroughput}(b), presents comparable PDCP throughput gains with respect to the HH option.

Finally, it is interesting to notice that, when the system implements a DC architecture for handover management, the traditional trade-off between latency and throughput no longer holds. In fact, despite the increased number of handover and switch events shown in Fig. \ref{fig:n_HO}(a), with respect to the baseline HH configuration, the UE experiences both a reduced latency and an increased PDCP throughput, thus enhancing the overall network quality of service.

\subsection{Variance Ratio}

In order to compare the variance of the rate experienced in time by a user, according to the different HO algorithms implemented (DC or HH, for fixed and dynamic TTT), we used the~ratio 
\beq
R_{\rm var} = \frac{\sigma_{{S}_{\rm PDCP}}}{\mathbb{E}[S_{\rm PDCP}]},
\label{eq:var}
\eeq
where $\mathbb{E}[{S}_{\rm PDCP}]$ is the mean value of the  PDCP throughput measured for each HO configuration and  $\sigma_{{S}_{\rm PDCP}}$ is its standard deviation, obtained over $N$ repetitions.
High values of $R_{\rm var} $ reflect remarkable channel instability, thus the rate would be affected by local variations and periodic degradations.

\begin{figure*}[t]
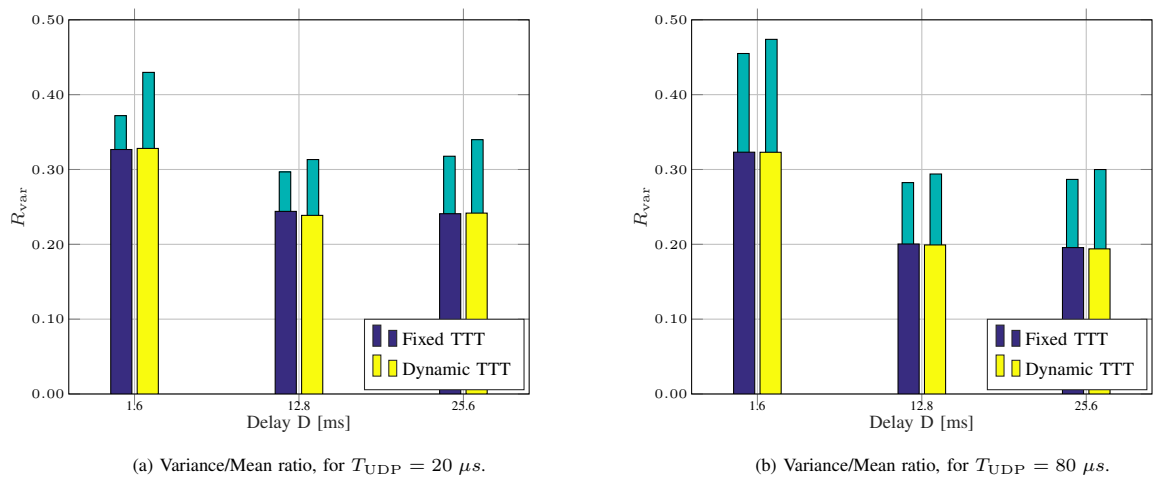

	\centering
	\begin{subfigure}[t]{0.45\textwidth}
		\setlength{\belowcaptionskip}{0cm}
		\iftikz
		\setlength{\belowcaptionskip}{0cm}
		\setlength\fwidth{0.75\textwidth}
		\setlength\fheight{0.61\textwidth}
		\input{./figures/thMeanStdRatio_10MB.tex}
		\else
		\includegraphics[width = \textwidth]{thMeanStdRatio_10MB}
		\fi
		\caption{Variance/Mean ratio,  for $T_{\rm UDP} = 20 \; \mu s$.}
	\end{subfigure}
	\begin{subfigure}[t]{0.45\textwidth}
		\setlength{\belowcaptionskip}{0cm}
		\iftikz
		\setlength{\belowcaptionskip}{0cm}
		\setlength\fwidth{0.75\textwidth}
		\setlength\fheight{0.61\textwidth}
		\input{./figures/thMeanStdRatio_10MB_80us.tex}
		\else
		\includegraphics[width = \textwidth]{thMeanStdRatio_10MB_80us}
		\fi
		\caption{Variance/Mean ratio,  for $T_{\rm UDP} = 80 \; \mu s$.} 
	\end{subfigure}
	\caption{Average ratio $R_{\rm var}$, for different values of the delay $D$ and the UDP packet interarrival time $T_{\rm UDP}$, for a fixed and dynamic TTT HO algorithm. Narrow bars refer to a hard handover configuration, while wide colored bars refer to a dual connectivity implementation. The RLC buffer size is $B_{\rm RLC}=10$ MB.}
	\label{fig:varianceRatio} 
\end{figure*}

Let $R_{\rm var,DC}$ and $R_{\rm var,HH}$ be the variance ratios of Equation \eqref{eq:var} for the fast switching with dual connectivity and hard handover configurations, respectively. 
From Fig. \ref{fig:varianceRatio}, we observe that $R_{\rm var,HH}$ is higher than $R_{\rm var,DC}$, for each value of the delay $D$, the HO metric and the UDP packet interarrival time $T_{\rm UDP}$, making it clear that the LTE eNB employed in a DC configuration can stabilize the rate, which is not subject to significant variations. 
In fact, in the portion of time in which the UE would experience zero gain if a hard handover architecture were implemented (due to an outage event), the rate would suffer a noticeable discrepancy with respect to the LOS values, thus increasing the rate variance throughout the simulation. This is not the case for the DC configuration, in which the UE can always be supported by the LTE eNB, even when a blockage event affects the scenario.
This result is fundamental for real-time applications, which require a long-term stable throughput to support high data rates and a consistently acceptable Quality of Experience for the users.

Furthermore, it can be seen that $R_{\rm var}$ increases when the CRT are collected more intensively.
In fact, even though reducing $D$ ensures better monitoring of the UE's motion and faster reaction to the channel variations (i.e., LOS/NLOS transitions or periodic modification of the small and large scale fading parameters of \textbf{H}), the user is affected by a higher number of handover and switch events, as depicted in Fig.  \ref{fig:n_HO}(a): in this way, the serving cell will be adapted regularly during the simulation, thereby causing large and periodic variation of the experienced throughput.
For the same reason, $R_{\rm var}$ is  higher when applying a dynamic TTT HO algorithm, since the handovers and switches outnumber those of a fixed TTT configuration.

Finally, to compare the DC and the HH architectures, we can consider the ratio
$R_{\rm DC/HH} = R_{\rm var,DC}/R_{\rm var,HH}$.
It assumes values lower than $1$, reflecting the lower variance of a DC configuration, with respect to the baseline HH option. 
We can therefore affirm that (i) $R_{\rm DC/HH}<1$ for every parameter combination and (ii) although the dynamic TTT HO approach shows an absolute higher variance than the fixed TTT one, the hard handover baseline suffers much more because of the aggressiveness of the dynamic TTT configuration than the DC architecture, and therefore $R_{\rm DC/HH,dyn} < R_{\rm DC/HH,fixed}$.

\subsection{RRC Traffic}

\begin{figure}[t]
	\centering
	\iftikz
	\setlength{\belowcaptionskip}{0cm}
	\setlength\fwidth{0.69\columnwidth}
	\setlength\fheight{0.56\columnwidth}
	\input{./figures/rrc_10MB.tex}
	\else
	\includegraphics[width = 0.45\textwidth]{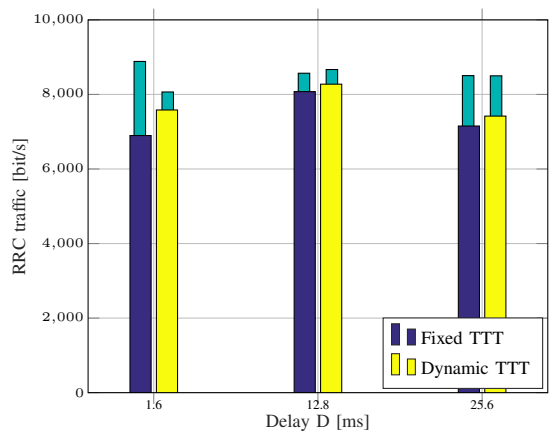}
	\fi
	\caption{Average amount of traffic at the RRC layer in bit/s, for different values of the delay $D$, for a fixed and dynamic TTT HO algorithm. Narrow bars refer to a hard handover configuration, while wide colored bars refer to a dual connectivity implementation. The RLC buffer size is $B_{\rm RLC}=10$ MB.}
	\label{fig:trafficRRC} 
\end{figure}

The RRC traffic is an indication of how many control operations are done by the UE-mmWave eNB pairs. Moreover, it is dependent also on the RRC PDU size\footnote{For example, a switch message contains $1$ byte for each of the bearers that should be switched, while an RRC connection reconfiguration message (which triggers the handover) carries several data structures, for a minimum of $59$ bytes for a single bearer reconfiguration.}. 

Fig. \ref{fig:trafficRRC} shows the RRC traffic for different values of the delay $D$. Notice that the RRC traffic is independent of the buffer size $B$, since even $10$ MB are enough to buffer the RRC PDUs, and of the UDP packet interarrival time $T_{\rm UDP}$.
It can be seen that fast switching causes an RRC traffic which is lower than for hard handover. 
The reason for this behavior is that, when implementing a DC solution, part of the control channel occupancy is due to the switches between the mmWave eNB and the LTE eNB, which use smaller control PDUs than standalone handover events with the HH architecture.
A lower RRC traffic is better, since it allows to allocate more resources to data transmission and, given the same amount of control overhead, it allows to scale to a larger number of users \cite{simutoolspolese}.

The RRC traffic is then higher for the dynamic TTT HO configuration due to the corresponding higher number of required handovers and switches shown in Fig. \ref{fig:n_HO}(a).

Finally, we highlight that the RRC traffic measured for a CRT intergeneration periodicity $D=1.6$ ms is lower than for $D\in\{12.8, 25.6\}$ ms, despite its higher number of required handovers and switches. The reason is that, when the CRT are very frequent, the UE is more intensively monitored, and can thus react more promptly when an outage or a channel update occurs. In this way, retransmissions of control PDUs are less probable and thus fewer messages need to be exchanged at the RRC layer.

\subsection{X2 Traffic}
\label{sec:x2}
\begin{figure*}[t]
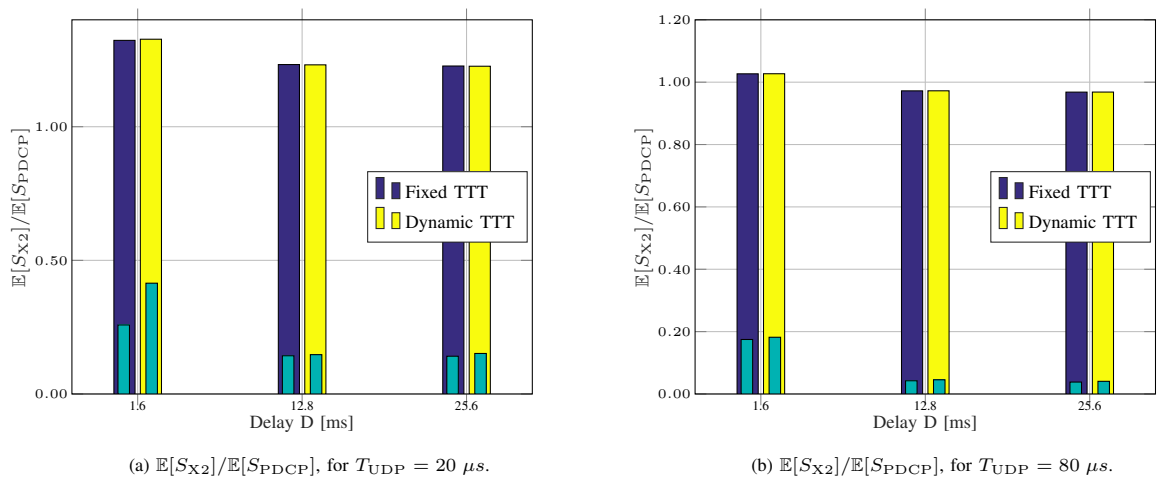

	\centering
	\begin{subfigure}[t]{0.45\textwidth}
		\setlength{\belowcaptionskip}{0cm}
		\iftikz
		\setlength{\belowcaptionskip}{0cm}
		\setlength\fwidth{0.75\textwidth}
		\setlength\fheight{0.61\textwidth}
		\input{./figures/x2_pdcp_ratio_10MB_20us.tex}
		\else
		\includegraphics[width = \textwidth]{x2_pdcp_ratio_10MB_20us}
		\fi
		\caption{$\mathbb{E}[S_{\rm X2}]/\mathbb{E}[S_{\rm PDCP}]$,  for $T_{\rm UDP} = 20 \; \mu s$.}
		\label{x2ratio_20}
	\end{subfigure}
	\begin{subfigure}[t]{0.45\textwidth}
		\setlength{\belowcaptionskip}{0cm}
		\iftikz
		\setlength{\belowcaptionskip}{0cm}
		\setlength\fwidth{0.75\textwidth}
		\setlength\fheight{0.61\textwidth}
		\input{./figures/x2_pdcp_ratio_10MB_80us.tex}
		\else
		\includegraphics[width = \textwidth]{x2_pdcp_ratio_10MB_80us}
		\fi
		\caption{$\mathbb{E}[S_{\rm X2}]/\mathbb{E}[S_{\rm PDCP}]$, for $T_{\rm UDP} = 80 \; \mu s$.}
		\label{x2ratio_80}
	\end{subfigure}
	\caption{Average ratio of X2 and PDCP throughput, for different values of the delay $D$ and of the UDP packet interarrival time $T_{\rm UDP}$, for a fixed and dynamic TTT HO algorithm. Narrow bars refer to a hard handover configuration, while wide colored bars refer to a dual connectivity implementation. The RLC buffer size is $B_{\rm RLC}=10$ MB.}
	\label{fig:x2ratio} 
\end{figure*}

\begin{figure*}[t]
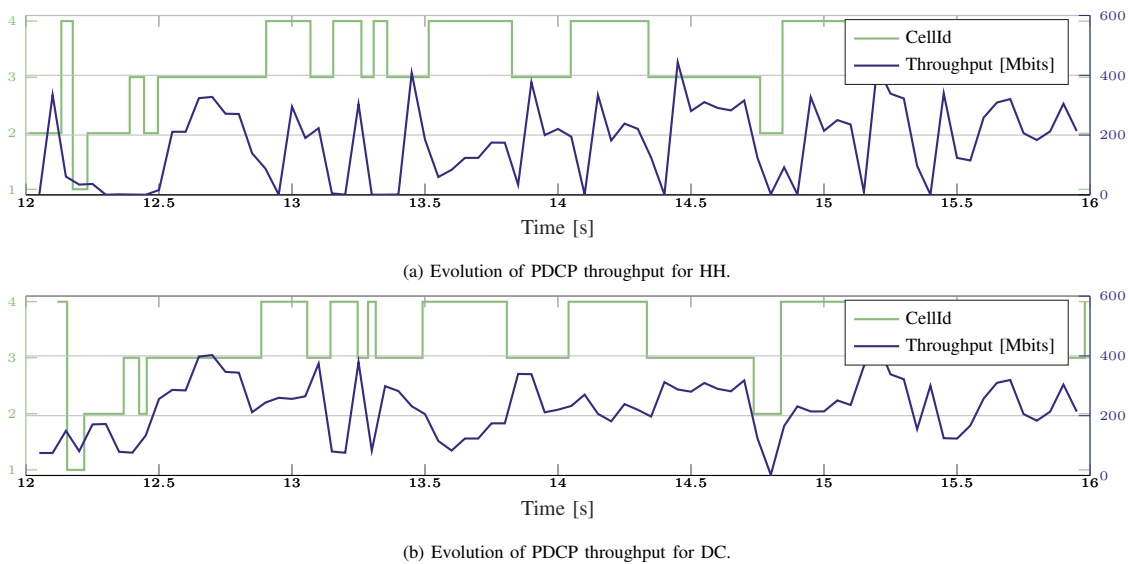

  \centering
  \begin{subfigure}[t]{\textwidth}
    \centering
    \setlength{\belowcaptionskip}{0cm}
    \iftikz
    \setlength\fwidth{0.78\textwidth}
    \setlength\fheight{0.17\textwidth}
    \input{figures/thInTime_B10_fixed150_1600_run15_hh.tex}    
    \else
    \includegraphics[width = \textwidth]{thInTime_B10_fixed150_1600_run15_hh}
    \fi
    \caption{Evolution of PDCP throughput for HH.}
  \end{subfigure}
  \begin{subfigure}[t]{\textwidth}
    \centering
    \setlength{\belowcaptionskip}{0cm}
    \iftikz
    \setlength\fwidth{0.78\textwidth}
    \setlength\fheight{0.17\textwidth}
    \input{figures/thInTime_B10_fixed150_1600_run15_dc.tex}
    \else
    \includegraphics[width = \textwidth]{thInTime_B10_fixed150_1600_run15_dc}
    \fi
    \caption{Evolution of PDCP throughput for DC.} 
  \end{subfigure}
  \caption{Evolution, for a specific simulation of duration $T_{\rm sim} = 20$ seconds, of the PDCP throughput and of the UE's instantaneous mmWave eNB association. We compare both the hard handover (above) and the dual connectivity (below) configurations, for the fixed TTT HO algorithm and a delay $D=1.6$ ms. The RLC buffer size is $B_{\rm RLC}=10$ MB. The green line represents the current cell over time, where cells from 2 to 4 are mmWave eNBs and cell 1 is the LTE eNB.}
  \label{fig:throughputEvolution} 
\end{figure*}

One drawback of the DC architecture is that it needs to forward PDCP PDUs from the LTE eNB to the mmWave eNB, besides forwarding the content of RLC buffers during switching and SCH events. On the other hand, the HH option only needs the second kind of forwarding during handovers. Therefore, the load on the X2 links connecting the different eNBs is lower for the HH solution, as  can be seen in Fig.~\ref{fig:x2ratio}, which shows the ratio between the average $\mathbb{E}[S_{\rm X2}]$ of the sum of the throughput $S_{\rm X2}$ in the six X2 links of the scenario and the average PDCP throughput $\mathbb{E}[S_{\rm PDCP}]$. It can be seen that for the DC architecture the ratio is close to 1, therefore the X2 links for such configuration must be dimensioned according to the target PDCP throughput for each mmWave eNB. For both  architectures the ratio is higher for the lower UDP interarrival time, since there are more packets buffered at the RLC layer that must be forwarded, and also for lower delay $D$, since there are more handover events. However, as we will discuss in more detail in Section~\ref{sec:finalComments}, the forwarding cost (in terms of inbound traffic to the mmWave eNB) of the DC architecture is similar to that of HH.

\begin{figure*}[t]
  \centering
  \begin{subfigure}[t]{0.5\textwidth}
    \setlength{\belowcaptionskip}{0cm}
    \centering
    \begin{tikzpicture}[font=\sffamily\scriptsize, scale=.85, every node/.style={scale=.85}]
      \centering
      \node[anchor=south west,inner sep=0] (image) at (0,0) {\includegraphics[width=\textwidth]{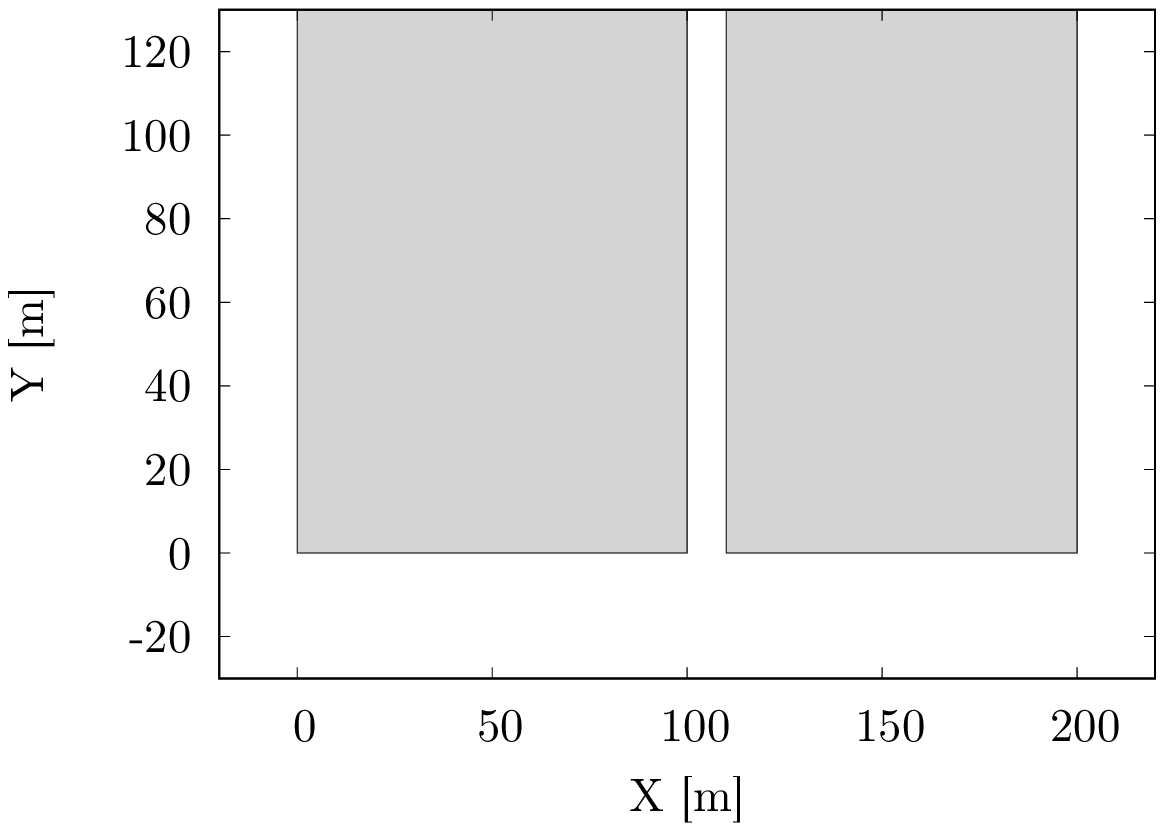}};
      \begin{scope}[x={(image.south east)},y={(image.north west)}]
      \filldraw[red,ultra thick] (0.248,0.3) circle (1.2pt);
      \node[anchor=west] at (0.252,0.3) (mm1label) {mmWave eNB};

      \filldraw[red,ultra thick] (0.87,0.3) circle (1.2pt);
      \node[anchor=east] at (0.87,0.3) (mm2label) {mmWave eNB};
      \filldraw[red,ultra thick] (0.57,0.83) circle (1.2pt);
      \node[anchor=south] at (0.57,0.835) (ltelabel) {LTE + mmWave eNB};

      \draw[line] (0.545, 0.31) -- node[anchor=north] {} (0.565, 0.31);
      \draw[sarrow] (0.565, 0.31) -- (0.565, 0.35);

      \filldraw[blue,ultra thick] (0.54, 0.31) circle (1.2pt);
      \node[anchor=north  ] at (0.54, 0.31) (ltelabel) {UE};
      \end{scope}   
    \end{tikzpicture}
    \caption{Corner case. The grey rectangles are buildings.}
    \label{fig:corner_case}
  \end{subfigure}
  \quad
  \begin{subfigure}[t]{0.45\textwidth}
    \setlength{\belowcaptionskip}{0cm}
    \iftikz
    \setlength{\belowcaptionskip}{0cm}
    \setlength\fwidth{0.75\textwidth}
    \setlength\fheight{0.61\textwidth}
    \input{./figures/latency_10MB_cornerCase_20us.tex}
    \else
    \includegraphics[width = \textwidth]{latency_10MB_cornerCase_20us}
    \fi
    \caption{Latency.}
  \end{subfigure}
  \caption{Average latency, for different values of the delay $D$ and for $T_{\rm UDP}=20\:\mu s$, comparing a fixed and dynamic TTT HO algorithm. The colored bars refer to a dual connectivity implementation for HO management. The RLC buffer size is $B=10$ MB and a \textit{corner scenario} is implemented, for a user moving at speed $v$.}
  \label{fig:TScenario} 
\end{figure*}

\subsection{Final Comments}
\label{sec:finalComments}

\textbf{Dual Connectivity vs. Hard Handover}: It can be seen that, in general, a multi-connectivity architecture performs better than the hard handover configuration.
The main benefit is the short time it takes to change radio access network and its enhancements are shown in terms of mainly: 
(i) \textit{latency}, which is reduced up to 50\% because the fast switching and SCH procedures are in general much faster than traditional handovers (although the number of SCH or switching events may be higher with DC), as observed in Fig.~\ref{fig:latency} and Fig.~\ref{fig:n_HO}(a); 
(ii) \textit{packet loss}, which is reduced since PDUs are less frequently buffered, thus reducing the overflow probability, as shown in Fig~\ref{fig:n_HO}(b). This is shown by the lower PDCP throughput of Fig.~\ref{fig:throughputEvolution}(a), referred to the HH configuration, with respect to that of the DC architecture of Fig.~\ref{fig:throughputEvolution}(b);
(iii) \textit{control signaling} related to the user plane which, despite an increase of the RRC traffic for the LTE eNB, is smaller with the DC solution (this allows the LTE eNB to handle the load of more UEs). This is supported by the results shown in Fig~\ref{fig:trafficRRC};
(iv) \textit{throughput variance}, where smaller rate variations are registered, with a reduction of $R_{\rm var}$ of up to 40\%, as observed in Fig.~\ref{fig:varianceRatio}. As an example, Fig.~\ref{fig:throughputEvolution}(a) shows periodic wide fluctuations of the throughput (which sometimes is even zero, when outages occur), while it settles on steady values when DC is applied, as in Fig.~\ref{fig:throughputEvolution}(b).

We also showed that, when the system implements the DC configuration, despite the increased number of handovers and switches, the UE can \textit{jointly} achieve both a reduced latency and an increased PDCP throughput, enhancing its overall quality of service. 
We have also examined the main cost of the DC architecture, showing in Section~\ref{sec:x2} that the X2 traffic for the DC option is higher than for the HH configuration because of the forwarding of packets from the LTE eNB to the mmWave ones. 
However, we must recall that, with the HH solution, the mmWave eNBs receive the packets from the core network through the S1 link, which is not used for the mmWave eNBs in the DC configuration. Therefore, when considering the overall inbound traffic to the mmWave eNBs on both the X2 and the S1 links, the costs of the two architectures may be equivalent. 
Given these considerations, we argue that the use of multi-connectivity for mobility management is to be preferred to the traditional hard handover approach.

\textbf{UDP interarrival time}: We observed that the general behaviors are similar for most metrics. However, the latency is much lower when $T_{\rm UDP}=80\:\mu s$, since RLC buffers are empty most of the time and fewer packets need to be forwarded during the switching and handover events. This justifies the  wider gap between DC and HH architectures, with respect to the $T_{\rm UDP}=20\:\mu$s case.

\textbf{CRT intergeneration delay and beamforming architecture}: We noticed remarkable differences between $D=1.6$ and $D=25.6$ ms (validating the choice of designing a \textit{digital} BF architecture, more complex but more efficient in terms of both latency and throughput) but almost no distinction between  the $D=12.8$ and $D=25.6$ ms configurations: we conclude that a \textit{hybrid} BF system at the mmWave eNB side is not to be preferred to an \textit{analog} one, since the complexity is increased while the overall performance is almost equivalent.

\textbf{Fixed vs. Dynamic TTT}: We showed that the second approach never results in a  performance degradation for any of the analyzed metrics. Moreover, we showed that it may also deliver tangible improvements in some specific scenarios where the traditional methods fail, such as the one shown in Fig. \ref{fig:TScenario}.
In this \textit{corner scenario}, the UE turns left at a T-junction and loses LOS with respect to both mmWave eNBs at the bottom. However, the mmWave eNB on top of the scenario is now in LOS, thus the handover should be triggered as quickly as possible.
From the result in Fig.~\ref{fig:TScenario}(b), we observe that in this case a dynamic and more aggressive approach is able to massively reduce latency compared to the fixed configuration, since a reduced TTT may be vital in this specific scheme, in which the user experiences a degraded rate until the handover to the LOS mmWave eNB is completed.
We indeed state that, since the dynamic TTT algorithm never underperforms the fixed TTT approach but is able to greatly improve the performance in specific scenarios, it should be preferred for  handover management.

\section{Conclusion And Future Work}
A limitation for the deployment of mmWave 5G systems is the rapidly changing dynamic channel caused by user mobility. The UE may be suddenly in outage with respect to all the mmWave eNBs, and a classic standalone architecture with traditional handovers cannot react quickly enough. In this paper we proposed a dual connectivity framework that, with the aid of a macro LTE eNB, can collect measurements and track the channel dynamics and perform fast switching to fall back to LTE and SCH for a fast handover among the mmWave eNBs. We showed, with an extensive simulation campaign, that the proposed framework is able to improve the performance of an end-to-end network with mmWave access links with respect to several metrics, including latency, throughput (in terms of both average and stability), radio control signaling and packet loss. Moreover, we presented and studied the performance of a dynamic TTT algorithm for SCH, showing that in some specific cases it may gain significantly with respect to a standard fixed TTT handover algorithm.

In our study, we focused on a simulated semi-statistical channel model with a realistic obstacles deployment to capture the mmWave dynamics. Due to the lack of temporally correlated mmWave channel measurements, it is currently not possible to develop an accurate analytical model for mobility-related scenarios, which on the other hand remains a very interesting and relevant item for future research.

\bibliographystyle{IEEEtran}
\bibliography{bibliography}

\end{document}

%% file: figures/handovers_10MB.tex
%
%
\definecolor{mycolor3}{rgb}{0.00000,0.70000,0.70000}%
\definecolor{mycolor2}{rgb}{0.97630,0.98310,0.05380}%
\definecolor{mycolor1}{rgb}{0.2196,0.1725,0.5020}%
\begin{tikzpicture}
\pgfplotsset{every tick label/.append style={font=\tiny}}
\pgfplotsset{scaled y ticks=false}

\begin{axis}[%
ybar=100pt,
width=\fwidth,
height=\fheight,
at={(0\fwidth,0\fheight)},
scale only axis,
bar shift auto,
xtick=data,
enlarge x limits=0.2,
bar width=8pt,
xticklabels={{1.6},{12.8},{25.6}},
xlabel style={font=\scriptsize\color{white!15!black}},
xlabel={Delay D [ms]},
ymin=0,
ymax=50,
ylabel style={font=\scriptsize\color{white!15!black}},
ylabel={\# HO events},
axis background/.style={fill=white},
xmajorgrids,
ymajorgrids,
ylabel shift = -5 pt,
xlabel shift = -5 pt,
yticklabel shift = -2 pt,
xticklabel shift = -5 pt,
yticklabel style={
            /pgf/number format/fixed,
            /pgf/number format/precision=0,
            /pgf/number format/fixed zerofill
        },
        scaled y ticks=false,
legend style={font=\scriptsize,at={(0.99,0.99)}, anchor=north east, legend cell align=left, align=left, draw=white!15!black}
]

\addplot[fill=mycolor1] coordinates {(1,35.9047619047619) (2,26.2333333333333) (3,21.3793103448276)};
\addplot[fill=mycolor2] coordinates {(1,42.0909090909091) (2,28.3666666666667) (3,23.4666666666667)};
\legend{Fixed TTT, Dynamic TTT}

\draw[fill=mycolor3] (0.88,0) rectangle (0.95,25.2857142857143);
\draw[fill=mycolor3] (1.05,0) rectangle (1.12,38.6666666666667);
\draw[fill=mycolor3] (1.88,0) rectangle (1.95,19.8333333333333);
\draw[fill=mycolor3] (2.05,0) rectangle (2.12,21.8333333333333);
\draw[fill=mycolor3] (2.88,0) rectangle (2.95,20.037037037037);
\draw[fill=mycolor3] (3.05,0) rectangle (3.12,21.1153846153846);




\end{axis}
\end{tikzpicture}%

%% file: figures/loss_10MB.tex
%
%
\definecolor{mycolor3}{rgb}{0.00000,0.70000,0.70000}%
\definecolor{mycolor2}{rgb}{0.97630,0.98310,0.05380}%
\definecolor{mycolor1}{rgb}{0.2196,0.1725,0.5020}%
\begin{tikzpicture}
\pgfplotsset{every tick label/.append style={font=\tiny}}
\pgfplotsset{scaled y ticks=false}

\begin{axis}[%
ybar=100pt,
width=\fwidth,
height=\fheight,
at={(0\fwidth,0\fheight)},
scale only axis,
bar shift auto,
xtick=data,
enlarge x limits=0.2,
bar width=8pt,
xticklabels={{1.6},{12.8},{25.6}},
xlabel style={font=\scriptsize\color{white!15!black}},
xlabel={Delay D [ms]},
ymin=0,
ymax=0.2,
ylabel style={font=\scriptsize\color{white!15!black}},
ylabel={$R_{\rm loss}$},
axis background/.style={fill=white},
xmajorgrids,
ymajorgrids,
ylabel shift = -5 pt,
xlabel shift = -5 pt,
yticklabel shift = -2 pt,
xticklabel shift = -5 pt,
yticklabel style={
            /pgf/number format/fixed,
            /pgf/number format/precision=2,
            /pgf/number format/fixed zerofill
        },
        scaled y ticks=false,
legend style={font=\scriptsize,at={(0.99,0.01)}, anchor=south east, legend cell align=left, align=left, draw=white!15!black}
]

\draw[fill=mycolor3] (0.89,0) rectangle (0.96,0.0945938095238095);
\draw[fill=mycolor3] (1.04,0) rectangle (1.11,0.145846);
\draw[fill=mycolor3] (1.89,0) rectangle (1.96,0.155061133333333);
\draw[fill=mycolor3] (2.04,0) rectangle (2.11,0.1545551);
\draw[fill=mycolor3] (2.89,0) rectangle (2.96,0.15715637037037);
\draw[fill=mycolor3] (3.04,0) rectangle (3.11,0.169573884615385);

\addplot[fill=mycolor1] coordinates {(1,0.0775357619047619) (2,0.144784) (3,0.145872275862069)};
\addplot[fill=mycolor2] coordinates {(1,0.0785844090909091) (2,0.1440639) (3,0.14465)};
\legend{Fixed TTT, Dynamic TTT}





\end{axis}
\end{tikzpicture}%

%% file: figures/latency_10MB.tex
%
%
\definecolor{mycolor3}{rgb}{0.00000,0.70000,0.70000}%
\definecolor{mycolor2}{rgb}{0.97630,0.98310,0.05380}%
\definecolor{mycolor1}{rgb}{0.2196,0.1725,0.5020}%
\begin{tikzpicture}
\pgfplotsset{every tick label/.append style={font=\tiny}}
\pgfplotsset{scaled y ticks=false}

\begin{axis}[%
ybar=100pt,
width=\fwidth,
height=\fheight,
at={(0\fwidth,0\fheight)},
scale only axis,
bar shift auto,
xtick=data,
enlarge x limits=0.2,
bar width=8pt,
xticklabels={{1.6},{12.8},{25.6}},
xlabel style={font=\scriptsize\color{white!15!black}},
xlabel={Delay D [ms]},
ymin=0,
ymax=0.15,
ylabel style={font=\scriptsize\color{white!15!black}},
ylabel={Latency [s]},
axis background/.style={fill=white},
xmajorgrids,
ymajorgrids,
ylabel shift = -5 pt,
xlabel shift = -5 pt,
yticklabel shift = -2 pt,
xticklabel shift = -5 pt,
yticklabel style={
            /pgf/number format/fixed,
            /pgf/number format/precision=2,
            /pgf/number format/fixed zerofill
        },
        scaled y ticks=false,
legend style={font=\scriptsize,at={(0.99,0.01)}, anchor=south east, legend cell align=left, align=left, draw=white!15!black}
]

\draw[fill=mycolor3] (0.89,0) rectangle (0.96,0.0861592834634151);
\draw[fill=mycolor3] (1.04,0) rectangle (1.11,0.129107505694966);
\draw[fill=mycolor3] (1.89,0) rectangle (1.96,0.134634444699187);
\draw[fill=mycolor3] (2.04,0) rectangle (2.11,0.134336487229916);
\draw[fill=mycolor3] (2.89,0) rectangle (2.96,0.136204782093668);
\draw[fill=mycolor3] (3.04,0) rectangle (3.11,0.147393491439155);

\addplot[fill=mycolor1] coordinates {(1,0.0796169194660547) (2,0.128151376828153) (3,0.128500033935813)};
\addplot[fill=mycolor2] coordinates {(1,0.0818070051773961) (2,0.127704291490167) (3,0.128321070243052)};
\legend{Fixed TTT, Dynamic TTT}





\end{axis}
\end{tikzpicture}%

%% file: figures/latency_10MB_80us.tex
%
%
\definecolor{mycolor3}{rgb}{0.00000,0.70000,0.70000}%
\definecolor{mycolor2}{rgb}{0.97630,0.98310,0.05380}%
\definecolor{mycolor1}{rgb}{0.2196,0.1725,0.5020}%
\begin{tikzpicture}
\pgfplotsset{every tick label/.append style={font=\tiny}}
\pgfplotsset{scaled y ticks=false}

\begin{axis}[%
ybar=100pt,
width=\fwidth,
height=\fheight,
at={(0\fwidth,0\fheight)},
scale only axis,
bar shift auto,
xtick=data,
enlarge x limits=0.2,
bar width=8pt,
xticklabels={{1.6},{12.8},{25.6}},
xlabel style={font=\scriptsize\color{white!15!black}},
xlabel={Delay D [ms]},
ymin=0,
ymax=0.015,
ylabel style={font=\scriptsize\color{white!15!black}},
ylabel={Latency [s]},
axis background/.style={fill=white},
xmajorgrids,
ymajorgrids,
ylabel shift = -5 pt,
xlabel shift = -5 pt,
yticklabel shift = -2 pt,
xticklabel shift = -5 pt,
yticklabel style={
            /pgf/number format/fixed,
            /pgf/number format/precision=3,
            /pgf/number format/fixed zerofill
        },
        scaled y ticks=false,
legend style={font=\scriptsize,at={(0.99,0.01)}, anchor=south east, legend cell align=left, align=left, draw=white!15!black}
]

\draw[fill=mycolor3] (0.89,0) rectangle (0.96,0.0142844580446827);
\draw[fill=mycolor3] (1.04,0) rectangle (1.11,0.0145701832756011);
\draw[fill=mycolor3] (1.89,0) rectangle (1.96,0.00768876868818329);
\draw[fill=mycolor3] (2.04,0) rectangle (2.11,0.00783738796757614);
\draw[fill=mycolor3] (2.89,0) rectangle (2.96,0.00718443518380387);
\draw[fill=mycolor3] (3.04,0) rectangle (3.11,0.00730748927247383);

\addplot[fill=mycolor1] coordinates {(1,0.00479081842406225) (2,0.00509291317390233) (3,0.00556471746530028)};
\addplot[fill=mycolor2] coordinates {(1,0.00476904229373129) (2,0.0050671370384712) (3,0.00551155364079505)};
\legend{Fixed TTT, Dynamic TTT}



\end{axis}
\end{tikzpicture}%

%% file: figures/th_10MB.tex
%
%
\definecolor{mycolor3}{rgb}{0.00000,0.70000,0.70000}%
\definecolor{mycolor2}{rgb}{0.97630,0.98310,0.05380}%
\definecolor{mycolor1}{rgb}{0.2196,0.1725,0.5020}%
\begin{tikzpicture}
\pgfplotsset{every tick label/.append style={font=\tiny}}
\pgfplotsset{scaled y ticks=false}

\begin{axis}[%
ybar=100pt,
width=\fwidth,
height=\fheight,
at={(0\fwidth,0\fheight)},
scale only axis,
bar shift auto,
xtick=data,
enlarge x limits=0.2,
bar width=8pt,
xticklabels={{1.6},{12.8},{25.6}},
xlabel style={font=\scriptsize\color{white!15!black}},
xlabel={Delay D [ms]},
ymin=0,
ymax=400,
ylabel style={font=\scriptsize\color{white!15!black}},
ylabel={PDCP Throughput [Mbit/s]},
axis background/.style={fill=white},
xmajorgrids,
ymajorgrids,
ylabel shift = -5 pt,
xlabel shift = -5 pt,
yticklabel shift = -2 pt,
xticklabel shift = -5 pt,
yticklabel style={
            /pgf/number format/fixed,
            /pgf/number format/precision=0,
            /pgf/number format/fixed zerofill
        },
        scaled y ticks=false,
legend style={font=\scriptsize,at={(0.99,0.01)}, anchor=south east, legend cell align=left, align=left, draw=white!15!black}
]

\addplot[fill=mycolor1] coordinates {(1,392.390416212399) (2,355.851288943985) (3,356.212091075934)};
\addplot[fill=mycolor2] coordinates {(1,392.211080950171) (2,356.493782320519) (3,356.755485342868)};
\legend{Fixed TTT, Dynamic TTT}

\draw[fill=mycolor3] (0.88,0) rectangle (0.95,383.880123142729);
\draw[fill=mycolor3] (1.05,0) rectangle (1.12,361.69884232026);
\draw[fill=mycolor3] (1.88,0) rectangle (1.95,350.948801362022);
\draw[fill=mycolor3] (2.05,0) rectangle (2.12,351.551340869779);
\draw[fill=mycolor3] (2.88,0) rectangle (2.95,350.417384068873);
\draw[fill=mycolor3] (3.05,0) rectangle (3.12,345.007767613626);




\end{axis}
\end{tikzpicture}%

%% file: figures/th_10MB_80us.tex
%
%
\definecolor{mycolor3}{rgb}{0.00000,0.70000,0.70000}%
\definecolor{mycolor2}{rgb}{0.97630,0.98310,0.05380}%
\definecolor{mycolor1}{rgb}{0.2196,0.1725,0.5020}%
\begin{tikzpicture}
\pgfplotsset{every tick label/.append style={font=\tiny}}
\pgfplotsset{scaled y ticks=false}

\begin{axis}[%
ybar=100pt,
width=\fwidth,
height=\fheight,
at={(0\fwidth,0\fheight)},
scale only axis,
bar shift auto,
xtick=data,
enlarge x limits=0.2,
bar width=8pt,
xticklabels={{1.6},{12.8},{25.6}},
xlabel style={font=\scriptsize\color{white!15!black}},
xlabel={Delay D [ms]},
ymin=0,
ymax=120,
ylabel style={font=\scriptsize\color{white!15!black}},
ylabel={PDCP Throughput [Mbit/s]},
axis background/.style={fill=white},
xmajorgrids,
ymajorgrids,
ylabel shift = -5 pt,
xlabel shift = -5 pt,
yticklabel shift = -2 pt,
xticklabel shift = -5 pt,
yticklabel style={
            /pgf/number format/fixed,
            /pgf/number format/precision=0,
            /pgf/number format/fixed zerofill
        },
        scaled y ticks=false,
legend style={font=\scriptsize,at={(0.99,0.01)}, anchor=south east, legend cell align=left, align=left, draw=white!15!black}
]

\addplot[fill=mycolor1] coordinates {(1,107.948023975755) (2,104.56166904412) (3,104.809783540888)};
\addplot[fill=mycolor2] coordinates {(1,108.057509067746) (2,104.645969002649) (3,104.889301881759)};
\legend{Fixed TTT, Dynamic TTT}

\draw[fill=mycolor3] (0.88,0) rectangle (0.95,106.772605493773);
\draw[fill=mycolor3] (1.05,0) rectangle (1.12,106.898056665947);
\draw[fill=mycolor3] (1.88,0) rectangle (1.95,104.219691239052);
\draw[fill=mycolor3] (2.05,0) rectangle (2.12,104.234918147628);
\draw[fill=mycolor3] (2.88,0) rectangle (2.95,104.292507618798);
\draw[fill=mycolor3] (3.05,0) rectangle (3.12,104.348404160275);




\end{axis}
\end{tikzpicture}%

%% file: figures/thMeanStdRatio_10MB.tex
%
%
\definecolor{mycolor3}{rgb}{0.00000,0.70000,0.70000}%
\definecolor{mycolor2}{rgb}{0.97630,0.98310,0.05380}%
\definecolor{mycolor1}{rgb}{0.2196,0.1725,0.5020}%
\begin{tikzpicture}
\pgfplotsset{every tick label/.append style={font=\tiny}}
\pgfplotsset{scaled y ticks=false}

\begin{axis}[%
ybar=100pt,
width=\fwidth,
height=\fheight,
at={(0\fwidth,0\fheight)},
scale only axis,
bar shift auto,
xtick=data,
enlarge x limits=0.2,
bar width=8pt,
xticklabels={{1.6},{12.8},{25.6}},
xlabel style={font=\scriptsize\color{white!15!black}},
xlabel={Delay D [ms]},
ymin=0,
ymax=0.5,
ylabel style={font=\scriptsize\color{white!15!black}},
ylabel={$R_{\rm var}$},
axis background/.style={fill=white},
xmajorgrids,
ymajorgrids,
ylabel shift = -5 pt,
xlabel shift = -5 pt,
yticklabel shift = -2 pt,
xticklabel shift = -5 pt,
yticklabel style={
            /pgf/number format/fixed,
            /pgf/number format/precision=2,
            /pgf/number format/fixed zerofill
        },
        scaled y ticks=false,
legend style={font=\scriptsize,at={(0.99,0.01)}, anchor=south east, legend cell align=left, align=left, draw=white!15!black}
]

\draw[fill=mycolor3] (0.88,0) rectangle (0.95,0.37194844223609);
\draw[fill=mycolor3] (1.05,0) rectangle (1.12,0.429776081804446);
\draw[fill=mycolor3] (1.88,0) rectangle (1.95,0.296746709483144);
\draw[fill=mycolor3] (2.05,0) rectangle (2.12,0.313272269426348);
\draw[fill=mycolor3] (2.88,0) rectangle (2.95,0.317724893975126);
\draw[fill=mycolor3] (3.05,0) rectangle (3.12,0.339828416593699);

\addplot[fill=mycolor1] coordinates {(1,0.326697889541045) (2,0.244019227693122) (3,0.240905110919684)};
\addplot[fill=mycolor2] coordinates {(1,0.328153863143912) (2,0.238678592219164) (3,0.241714470834906)};
\legend{Fixed TTT, Dynamic TTT}


\end{axis}
\end{tikzpicture}%

%% file: figures/thMeanStdRatio_10MB_80us.tex
%
%
\definecolor{mycolor3}{rgb}{0.00000,0.70000,0.70000}%
\definecolor{mycolor2}{rgb}{0.97630,0.98310,0.05380}%
\definecolor{mycolor1}{rgb}{0.2196,0.1725,0.5020}%
\begin{tikzpicture}
\pgfplotsset{every tick label/.append style={font=\tiny}}
\pgfplotsset{scaled y ticks=false}

\begin{axis}[%
ybar=100pt,
width=\fwidth,
height=\fheight,
at={(0\fwidth,0\fheight)},
scale only axis,
bar shift auto,
xtick=data,
enlarge x limits=0.2,
bar width=8pt,
xticklabels={{1.6},{12.8},{25.6}},
xlabel style={font=\scriptsize\color{white!15!black}},
xlabel={Delay D [ms]},
ymin=0,
ymax=0.5,
ylabel style={font=\scriptsize\color{white!15!black}},
ylabel={$R_{\rm var}$},
axis background/.style={fill=white},
xmajorgrids,
ymajorgrids,
ylabel shift = -5 pt,
xlabel shift = -5 pt,
yticklabel shift = -2 pt,
xticklabel shift = -5 pt,
yticklabel style={
            /pgf/number format/fixed,
            /pgf/number format/precision=2,
            /pgf/number format/fixed zerofill
        },
        scaled y ticks=false,
legend style={font=\scriptsize,at={(0.99,0.01)}, anchor=south east, legend cell align=left, align=left, draw=white!15!black}
]

\draw[fill=mycolor3] (0.88,0) rectangle (0.95,0.454994899098121);
\draw[fill=mycolor3] (1.05,0) rectangle (1.12,0.473907869625518);
\draw[fill=mycolor3] (1.88,0) rectangle (1.95,0.282453523142037);
\draw[fill=mycolor3] (2.05,0) rectangle (2.12,0.293967228837309);
\draw[fill=mycolor3] (2.88,0) rectangle (2.95,0.286753542308761);
\draw[fill=mycolor3] (3.05,0) rectangle (3.12,0.299940502003315);

\addplot[fill=mycolor1] coordinates {(1,0.323110386507068) (2,0.20043852779469) (3,0.195622792180212)};
\addplot[fill=mycolor2] coordinates {(1,0.323007241668221) (2,0.199167876839701) (3,0.193921585978347)};
\legend{Fixed TTT, Dynamic TTT}



\end{axis}
\end{tikzpicture}%

%% file: figures/rrc_10MB.tex
%
%
\definecolor{mycolor3}{rgb}{0.00000,0.70000,0.70000}%
\definecolor{mycolor2}{rgb}{0.97630,0.98310,0.05380}%
\definecolor{mycolor1}{rgb}{0.2196,0.1725,0.5020}%
\begin{tikzpicture}
\pgfplotsset{every tick label/.append style={font=\tiny}}
\pgfplotsset{scaled y ticks=false}

\begin{axis}[%
ybar=100pt,
width=\fwidth,
height=\fheight,
at={(0\fwidth,0\fheight)},
scale only axis,
bar shift auto,
xtick=data,
enlarge x limits=0.2,
bar width=8pt,
xticklabels={{1.6},{12.8},{25.6}},
xlabel style={font=\scriptsize\color{white!15!black}},
xlabel={Delay D [ms]},
ymin=0,
ymax=10000,
ylabel style={font=\scriptsize\color{white!15!black}},
ylabel={RRC traffic [bit/s]},
axis background/.style={fill=white},
xmajorgrids,
ymajorgrids,
ylabel shift = -5 pt,
xlabel shift = -5 pt,
yticklabel shift = -2 pt,
xticklabel shift = -5 pt,
yticklabel style={
            /pgf/number format/fixed,
            /pgf/number format/precision=0,
            /pgf/number format/fixed zerofill
        },
        scaled y ticks=false,
legend style={font=\scriptsize,at={(0.99,0.01)}, anchor=south east, legend cell align=left, align=left, draw=white!15!black}
]

\draw[fill=mycolor3] (0.88,0) rectangle (0.95,8885.10824066325);
\draw[fill=mycolor3] (1.05,0) rectangle (1.12,8066.05200143961);
\draw[fill=mycolor3] (1.88,0) rectangle (1.95,8569.11437600234);
\draw[fill=mycolor3] (2.05,0) rectangle (2.12,8667.83306735709);
\draw[fill=mycolor3] (2.88,0) rectangle (2.95,8503.66343349659);
\draw[fill=mycolor3] (3.05,0) rectangle (3.12,8499.47816280954);

\addplot[fill=mycolor1] coordinates {(1,6897.36710895706) (2,8075.17903930555) (3,7152.61349259162)};
\addplot[fill=mycolor2] coordinates {(1,7582.67325161909) (2,8275.14257206908) (3,7418.91996279457)};
\legend{Fixed TTT, Dynamic TTT}


\end{axis}
\end{tikzpicture}%

%% file: figures/x2_pdcp_ratio_10MB_20us.tex
%
%
\definecolor{mycolor3}{rgb}{0.00000,0.70000,0.70000}%
\definecolor{mycolor2}{rgb}{0.97630,0.98310,0.05380}%
\definecolor{mycolor1}{rgb}{0.2196,0.1725,0.5020}%
\begin{tikzpicture}
\pgfplotsset{every tick label/.append style={font=\tiny}}
\pgfplotsset{scaled y ticks=false}

\begin{axis}[%
ybar=100pt,
width=\fwidth,
height=\fheight,
at={(0\fwidth,0\fheight)},
scale only axis,
bar shift auto,
xtick=data,
enlarge x limits=0.2,
bar width=8pt,
xticklabels={{1.6},{12.8},{25.6}},
xlabel style={font=\scriptsize\color{white!15!black}},
xlabel={Delay D [ms]},
ymin=0,
ymax=1.4,
ylabel style={font=\scriptsize\color{white!15!black}},
ylabel={$\mathbb{E}[S_{\rm X2}]/\mathbb{E}[S_{\rm PDCP}]$},
axis background/.style={fill=white},
xmajorgrids,
ymajorgrids,
ylabel shift = -5 pt,
xlabel shift = -5 pt,
yticklabel shift = -2 pt,
xticklabel shift = -5 pt,
yticklabel style={
            /pgf/number format/fixed,
            /pgf/number format/precision=2,
            /pgf/number format/fixed zerofill
        },
        scaled y ticks=false,
legend style={font=\scriptsize,at={(0.99,0.5)}, anchor=east, legend cell align=left, align=left, draw=white!15!black}
]

\addplot[fill=mycolor1] coordinates {(1,1.32312146118084) (2,1.2326216522262) (3,1.22718336804434)};
\addplot[fill=mycolor2] coordinates {(1,1.32741401415678) (2,1.23152293231334) (3,1.22634710128333)};
\legend{Fixed TTT, Dynamic TTT}

\draw[fill=mycolor3] (0.88,0) rectangle (0.95,0.25829834359255);
\draw[fill=mycolor3] (1.05,0) rectangle (1.12,0.414564587038033);
\draw[fill=mycolor3] (1.88,0) rectangle (1.95,0.143034539678475);
\draw[fill=mycolor3] (2.05,0) rectangle (2.12,0.14713173435738);
\draw[fill=mycolor3] (2.88,0) rectangle (2.95,0.141255991868509);
\draw[fill=mycolor3] (3.05,0) rectangle (3.12,0.151633095526436);




\end{axis}
\end{tikzpicture}%

%% file: figures/x2_pdcp_ratio_10MB_80us.tex
%
%
\definecolor{mycolor3}{rgb}{0.00000,0.70000,0.70000}%
\definecolor{mycolor2}{rgb}{0.97630,0.98310,0.05380}%
\definecolor{mycolor1}{rgb}{0.2196,0.1725,0.5020}%
\begin{tikzpicture}
\pgfplotsset{every tick label/.append style={font=\tiny}}
\pgfplotsset{scaled y ticks=false}

\begin{axis}[%
ybar=100pt,
width=\fwidth,
height=\fheight,
at={(0\fwidth,0\fheight)},
scale only axis,
bar shift auto,
xtick=data,
enlarge x limits=0.2,
bar width=8pt,
xticklabels={{1.6},{12.8},{25.6}},
xlabel style={font=\scriptsize\color{white!15!black}},
xlabel={Delay D [ms]},
ymin=0,
ymax=1.2,
ylabel style={font=\scriptsize\color{white!15!black}},
ylabel={$\mathbb{E}[S_{\rm X2}]/\mathbb{E}[S_{\rm PDCP}]$},
axis background/.style={fill=white},
xmajorgrids,
ymajorgrids,
ylabel shift = -5 pt,
xlabel shift = -5 pt,
yticklabel shift = -2 pt,
xticklabel shift = -5 pt,
yticklabel style={
            /pgf/number format/fixed,
            /pgf/number format/precision=2,
            /pgf/number format/fixed zerofill
        },
        scaled y ticks=false,
legend style={font=\scriptsize,at={(0.99,0.5)}, anchor=east, legend cell align=left, align=left, draw=white!15!black}
]

\addplot[fill=mycolor1] coordinates {(1,1.02680439215631) (2,0.972173788716325) (3,0.9681417937681)};
\addplot[fill=mycolor2] coordinates {(1,1.02705403842874) (2,0.972347944340673) (3,0.968295023567313)};
\legend{Fixed TTT, Dynamic TTT}

\draw[fill=mycolor3] (0.88,0) rectangle (0.95,0.174788954147193);
\draw[fill=mycolor3] (1.05,0) rectangle (1.12,0.181954715490663);
\draw[fill=mycolor3] (1.88,0) rectangle (1.95,0.0424242112031599);
\draw[fill=mycolor3] (2.05,0) rectangle (2.12,0.0455543380182443);
\draw[fill=mycolor3] (2.88,0) rectangle (2.95,0.0382012674598953);
\draw[fill=mycolor3] (3.05,0) rectangle (3.12,0.0406200661621887);




\end{axis}
\end{tikzpicture}%

%% file: figures/thInTime_B10_fixed150_1600_run15_hh.tex
%
%
\definecolor{mycolor1}{rgb}{0.53003,0.74911,0.46611}%
\definecolor{mycolor2}{rgb}{0.20810,0.16630,0.52920}%
\begin{tikzpicture}
\tikzstyle{every node}=[font=\scriptsize]
\pgfplotsset{every x tick label/.append style={font=\tiny, yshift=0.5ex}}
\pgfplotsset{every y tick label/.append style={font=\tiny}}

\begin{axis}[%
width=\fwidth,
height=0.774\fheight,
at={(-0\fwidth,-0\fheight)},
scale only axis,
xmin=12,
xmax=16,
xlabel style={font=\color{white!15!black}\footnotesize},
xlabel shift = -2 pt,
xlabel={Time [s]},
separate axis lines,
every outer y axis line/.append style={mycolor1},
every y tick label/.append style={font=\color{mycolor1}\tiny},
every y tick/.append style={mycolor1},
ymin=0.9,
ymax=4.1,
ytick={1, 2, 3, 4},
axis background/.style={fill=white},
]
\addplot[const plot, color=mycolor1, line width=0.8pt, forget plot] table[row sep=crcr] {%
12.0123	2\\
12.1332	4\\
12.1762	1\\
12.2313	2\\
12.3903	3\\
12.4443	2\\
12.4963	3\\
12.9033	4\\
13.0698	3\\
13.1553	4\\
13.2613	3\\
13.3073	4\\
13.3582	3\\
13.5143	4\\
13.8273	3\\
14.0483	4\\
14.3403	3\\
14.7603	2\\
14.8443	4\\
15.1233	3\\
15.3563	4\\
15.5382	3\\
};
\label{plotyyref:leg1}

\end{axis}

\begin{axis}[%
width=\fwidth,
height=0.774\fheight,
at={(-0\fwidth,-0\fheight)},
scale only axis,
xmin=12,
xmax=16,
every outer y axis line/.append style={mycolor2},
every y tick label/.append style={font=\color{mycolor2}\tiny},
every y tick/.append style={mycolor2},
ymajorgrids,
ymin=0,
ymax=600,
ytick={  0, 200, 400, 600},
axis x line*=bottom,
axis y line*=right,
legend style={legend cell align=left, align=left, draw=white!15!black}
]
\addlegendimage{/pgfplots/refstyle=plotyyref:leg1}
\addlegendentry{CellId}
\addplot [color=mycolor2, line width=0.8pt]
  table[row sep=crcr]{%
12.05	0.67952\\
12.1	334.24448\\
12.15	61.05568\\
12.2	34.24832\\
12.25	36.44048\\
12.3	0.34224\\
12.35	1.6864\\
12.4	0.8512\\
12.45	0.35024\\
12.5	16.03376\\
12.55	210.9736\\
12.6	210.96864\\
12.65	323.45152\\
12.7	327.49888\\
12.75	271.34176\\
12.8	270.49856\\
12.85	138.95936\\
12.9000000000001	87.86944\\
12.9500000000001	0.67952\\
13.0000000000001	294.95136\\
13.0500000000001	190.22592\\
13.1000000000001	222.7864\\
13.1500000000001	4.89856\\
13.2000000000001	0.67952\\
13.2500000000001	303.56\\
13.3000000000001	0.35024\\
13.3500000000001	0.51888\\
13.4000000000001	1.0168\\
13.4500000000001	408.1088\\
13.5000000000001	184.83744\\
13.5500000000001	59.70352\\
13.6000000000001	83.64544\\
13.6500000000001	123.61312\\
13.7000000000001	123.61312\\
13.7500000000001	174.87968\\
13.8000000000001	174.37376\\
13.8500000000001	33.40368\\
13.9000000000001	376.74176\\
13.9500000000001	199.8384\\
14.0000000000001	220.41248\\
14.0500000000001	194.28128\\
14.1000000000001	0.84816\\
14.1500000000001	334.07584\\
14.2000000000001	181.79392\\
14.2500000000001	238.28832\\
14.3000000000001	220.58112\\
14.3500000000001	124.63296\\
14.4000000000001	0.67952\\
14.4500000000001	443.69184\\
14.5000000000001	280.27968\\
14.5500000000001	310.12896\\
14.6000000000001	290.73536\\
14.6500000000001	282.472\\
14.7000000000001	315.69408\\
14.7500000000001	123.28384\\
14.8000000000001	2.36096\\
14.8500000000001	91.92176\\
14.9000000000001	1.18544\\
14.9500000000001	326.99296\\
15.0000000000001	214.1728\\
15.0500000000001	250.09312\\
15.1000000000001	235.2528\\
15.1500000000001	9.28816\\
15.2000000000001	429.52608\\
15.2500000000001	338.46048\\
15.3000000000001	322.27104\\
15.3500000000001	96.80736\\
15.4000000000001	0.67952\\
15.4500000000001	339.64096\\
15.5000000000001	123.61312\\
15.5500000000001	115.02048\\
15.6000000000001	258.36144\\
15.6500000000001	308.77984\\
15.7000000000001	320.416\\
15.7500000000001	206.24672\\
15.8000000000001	183.31168\\
15.8500000000001	212.31776\\
15.9000000000001	304.73248\\
15.9500000000001	213.49824\\
};
\addlegendentry{Throughput [Mbits]}

\end{axis}
\end{tikzpicture}%

%% file: figures/thInTime_B10_fixed150_1600_run15_dc.tex
%
%
\definecolor{mycolor1}{rgb}{0.53003,0.74911,0.46611}%
\definecolor{mycolor2}{rgb}{0.20784,0.16471,0.52941}%
\definecolor{mycolor3}{rgb}{0.20810,0.16630,0.52920}%
\begin{tikzpicture}
\tikzstyle{every node}=[font=\scriptsize]
\pgfplotsset{every x tick label/.append style={font=\tiny, yshift=0.5ex}}
\pgfplotsset{every y tick label/.append style={font=\tiny}}

\begin{axis}[%
width=\fwidth,
height=0.774\fheight,
at={(0\fwidth,0\fheight)},
scale only axis,
xmin=12,
xmax=16,
xlabel style={font=\color{white!15!black}\footnotesize},
xlabel shift = -2 pt,
xlabel={Time [s]},
separate axis lines,
every outer y axis line/.append style={mycolor1},
every y tick label/.append style={font=\color{mycolor1}\tiny},
every y tick/.append style={mycolor1},
ymin=0.9,
ymax=4.1,
ytick={1, 2, 3, 4},
axis background/.style={fill=white},
]
\addplot[const plot, color=mycolor1, line width=0.8pt, forget plot] table[row sep=crcr] {%
12.1184	4\\
12.1184	4\\
12.1184	4\\
12.1184	4\\
12.1184	4\\
12.1184	4\\
12.1184	4\\
12.1184	4\\
12.1184	4\\
12.1184	4\\
12.1184	4\\
12.1184	4\\
12.1184	4\\
12.1184	4\\
12.1184	4\\
12.1184	4\\
12.1184	4\\
12.1184	4\\
12.1184	4\\
12.1184	4\\
12.1184	4\\
12.1184	4\\
12.1184	4\\
12.1184	4\\
12.1184	4\\
12.1184	4\\
12.1184	4\\
12.1184	4\\
12.1184	4\\
12.1184	4\\
12.1184	4\\
12.1184	4\\
12.1184	4\\
12.1184	4\\
12.1184	4\\
12.1184	4\\
12.1184	4\\
12.1184	4\\
12.1184	4\\
12.1184	4\\
12.1184	4\\
12.1184	4\\
12.1184	4\\
12.155	1\\
12.2192	2\\
12.2192	2\\
12.2192	2\\
12.2192	2\\
12.2192	2\\
12.2192	2\\
12.2192	2\\
12.2192	2\\
12.2192	2\\
12.2192	2\\
12.2192	2\\
12.2192	2\\
12.2192	2\\
12.2192	2\\
12.2192	2\\
12.2192	2\\
12.2192	2\\
12.2192	2\\
12.2192	2\\
12.2192	2\\
12.2192	2\\
12.2192	2\\
12.2192	2\\
12.2192	2\\
12.2192	2\\
12.2192	2\\
12.2192	2\\
12.2192	2\\
12.2192	2\\
12.2192	2\\
12.2192	2\\
12.2192	2\\
12.2192	2\\
12.2192	2\\
12.2192	2\\
12.2192	2\\
12.2192	2\\
12.2192	2\\
12.2192	2\\
12.2192	2\\
12.2192	2\\
12.2192	2\\
12.2192	2\\
12.2192	2\\
12.3678	3\\
12.3678	3\\
12.3678	3\\
12.3678	3\\
12.3678	3\\
12.3678	3\\
12.3678	3\\
12.3678	3\\
12.3678	3\\
12.3678	3\\
12.3678	3\\
12.3678	3\\
12.3678	3\\
12.3678	3\\
12.3678	3\\
12.3678	3\\
12.3678	3\\
12.3678	3\\
12.3678	3\\
12.3678	3\\
12.3678	3\\
12.3678	3\\
12.3678	3\\
12.3678	3\\
12.3678	3\\
12.3678	3\\
12.3678	3\\
12.3678	3\\
12.3678	3\\
12.3678	3\\
12.3678	3\\
12.3678	3\\
12.3678	3\\
12.3678	3\\
12.3678	3\\
12.3678	3\\
12.3678	3\\
12.3678	3\\
12.3678	3\\
12.3678	3\\
12.3678	3\\
12.3678	3\\
12.3678	3\\
12.3678	3\\
12.3678	3\\
12.4256	2\\
12.4256	2\\
12.4256	2\\
12.4256	2\\
12.4256	2\\
12.4256	2\\
12.4256	2\\
12.4256	2\\
12.4256	2\\
12.4256	2\\
12.4256	2\\
12.4256	2\\
12.4256	2\\
12.4256	2\\
12.4256	2\\
12.4256	2\\
12.4256	2\\
12.4256	2\\
12.4256	2\\
12.4256	2\\
12.4256	2\\
12.4256	2\\
12.4256	2\\
12.4256	2\\
12.4256	2\\
12.4256	2\\
12.4256	2\\
12.4256	2\\
12.4256	2\\
12.4256	2\\
12.4256	2\\
12.4256	2\\
12.4256	2\\
12.4256	2\\
12.4256	2\\
12.4256	2\\
12.4256	2\\
12.4256	2\\
12.4256	2\\
12.4256	2\\
12.4256	2\\
12.4256	2\\
12.4256	2\\
12.4256	2\\
12.4256	2\\
12.4256	2\\
12.4544	3\\
12.4544	3\\
12.4544	3\\
12.4544	3\\
12.4544	3\\
12.4544	3\\
12.4544	3\\
12.4544	3\\
12.4544	3\\
12.4544	3\\
12.4544	3\\
12.4544	3\\
12.4544	3\\
12.4544	3\\
12.4544	3\\
12.4544	3\\
12.4544	3\\
12.4544	3\\
12.4544	3\\
12.4544	3\\
12.4544	3\\
12.4544	3\\
12.4544	3\\
12.4544	3\\
12.4544	3\\
12.4544	3\\
12.4544	3\\
12.4544	3\\
12.4544	3\\
12.4544	3\\
12.4544	3\\
12.4544	3\\
12.4544	3\\
12.4544	3\\
12.4544	3\\
12.4544	3\\
12.4544	3\\
12.4544	3\\
12.4544	3\\
12.4544	3\\
12.4544	3\\
12.4544	3\\
12.4544	3\\
12.4544	3\\
12.4544	3\\
12.4544	3\\
12.4544	3\\
12.8848	4\\
12.8848	4\\
12.8848	4\\
12.8848	4\\
12.8848	4\\
12.8848	4\\
12.8848	4\\
12.8848	4\\
12.8848	4\\
12.8848	4\\
12.8848	4\\
12.8848	4\\
12.8848	4\\
12.8848	4\\
12.8848	4\\
12.8848	4\\
12.8848	4\\
12.8848	4\\
12.8848	4\\
12.8848	4\\
12.8848	4\\
12.8848	4\\
12.8848	4\\
12.8848	4\\
12.8848	4\\
12.8848	4\\
12.8848	4\\
12.8848	4\\
12.8848	4\\
12.8848	4\\
12.8848	4\\
12.8848	4\\
12.8848	4\\
12.8848	4\\
12.8848	4\\
12.8848	4\\
12.8848	4\\
12.8848	4\\
12.8848	4\\
12.8848	4\\
12.8848	4\\
12.8848	4\\
12.8848	4\\
12.8848	4\\
12.8848	4\\
12.8848	4\\
12.8848	4\\
12.8848	4\\
13.0576	3\\
13.0576	3\\
13.0576	3\\
13.0576	3\\
13.0576	3\\
13.0576	3\\
13.0576	3\\
13.0576	3\\
13.0576	3\\
13.0576	3\\
13.0576	3\\
13.0576	3\\
13.0576	3\\
13.0576	3\\
13.0576	3\\
13.0576	3\\
13.0576	3\\
13.0576	3\\
13.0576	3\\
13.0576	3\\
13.0576	3\\
13.0576	3\\
13.0576	3\\
13.0576	3\\
13.0576	3\\
13.0576	3\\
13.0576	3\\
13.0576	3\\
13.0576	3\\
13.0576	3\\
13.0576	3\\
13.0576	3\\
13.0576	3\\
13.0576	3\\
13.0576	3\\
13.0576	3\\
13.0576	3\\
13.0576	3\\
13.0576	3\\
13.0576	3\\
13.0576	3\\
13.0576	3\\
13.0576	3\\
13.0576	3\\
13.0576	3\\
13.0576	3\\
13.0576	3\\
13.0576	3\\
13.0576	3\\
13.145	4\\
13.145	4\\
13.145	4\\
13.145	4\\
13.145	4\\
13.145	4\\
13.145	4\\
13.145	4\\
13.145	4\\
13.145	4\\
13.145	4\\
13.145	4\\
13.145	4\\
13.145	4\\
13.145	4\\
13.145	4\\
13.145	4\\
13.145	4\\
13.145	4\\
13.145	4\\
13.145	4\\
13.145	4\\
13.145	4\\
13.145	4\\
13.145	4\\
13.145	4\\
13.145	4\\
13.145	4\\
13.145	4\\
13.145	4\\
13.145	4\\
13.145	4\\
13.145	4\\
13.145	4\\
13.145	4\\
13.145	4\\
13.145	4\\
13.145	4\\
13.145	4\\
13.145	4\\
13.145	4\\
13.145	4\\
13.145	4\\
13.145	4\\
13.145	4\\
13.145	4\\
13.145	4\\
13.145	4\\
13.145	4\\
13.145	4\\
13.2472	3\\
13.2472	3\\
13.2472	3\\
13.2472	3\\
13.2472	3\\
13.2472	3\\
13.2472	3\\
13.2472	3\\
13.2472	3\\
13.2472	3\\
13.2472	3\\
13.2472	3\\
13.2472	3\\
13.2472	3\\
13.2472	3\\
13.2472	3\\
13.2472	3\\
13.2472	3\\
13.2472	3\\
13.2472	3\\
13.2472	3\\
13.2472	3\\
13.2472	3\\
13.2472	3\\
13.2472	3\\
13.2472	3\\
13.2472	3\\
13.2472	3\\
13.2472	3\\
13.2472	3\\
13.2472	3\\
13.2472	3\\
13.2472	3\\
13.2472	3\\
13.2472	3\\
13.2472	3\\
13.2472	3\\
13.2472	3\\
13.2472	3\\
13.2472	3\\
13.2472	3\\
13.2472	3\\
13.2472	3\\
13.2472	3\\
13.2472	3\\
13.2472	3\\
13.2472	3\\
13.2472	3\\
13.2472	3\\
13.2472	3\\
13.2472	3\\
13.2864	4\\
13.2864	4\\
13.2864	4\\
13.2864	4\\
13.2864	4\\
13.2864	4\\
13.2864	4\\
13.2864	4\\
13.2864	4\\
13.2864	4\\
13.2864	4\\
13.2864	4\\
13.2864	4\\
13.2864	4\\
13.2864	4\\
13.2864	4\\
13.2864	4\\
13.2864	4\\
13.2864	4\\
13.2864	4\\
13.2864	4\\
13.2864	4\\
13.2864	4\\
13.2864	4\\
13.2864	4\\
13.2864	4\\
13.2864	4\\
13.2864	4\\
13.2864	4\\
13.2864	4\\
13.2864	4\\
13.2864	4\\
13.2864	4\\
13.2864	4\\
13.2864	4\\
13.2864	4\\
13.2864	4\\
13.2864	4\\
13.2864	4\\
13.2864	4\\
13.2864	4\\
13.2864	4\\
13.2864	4\\
13.2864	4\\
13.2864	4\\
13.2864	4\\
13.2864	4\\
13.2864	4\\
13.2864	4\\
13.2864	4\\
13.2864	4\\
13.2864	4\\
13.3152	3\\
13.3152	3\\
13.3152	3\\
13.3152	3\\
13.3152	3\\
13.3152	3\\
13.3152	3\\
13.3152	3\\
13.3152	3\\
13.3152	3\\
13.3152	3\\
13.3152	3\\
13.3152	3\\
13.3152	3\\
13.3152	3\\
13.3152	3\\
13.3152	3\\
13.3152	3\\
13.3152	3\\
13.3152	3\\
13.3152	3\\
13.3152	3\\
13.3152	3\\
13.3152	3\\
13.3152	3\\
13.3152	3\\
13.3152	3\\
13.3152	3\\
13.3152	3\\
13.3152	3\\
13.3152	3\\
13.3152	3\\
13.3152	3\\
13.3152	3\\
13.3152	3\\
13.3152	3\\
13.3152	3\\
13.3152	3\\
13.3152	3\\
13.3152	3\\
13.3152	3\\
13.3152	3\\
13.3152	3\\
13.3152	3\\
13.3152	3\\
13.3152	3\\
13.3152	3\\
13.3152	3\\
13.3152	3\\
13.3152	3\\
13.3152	3\\
13.3152	3\\
13.3152	3\\
13.4912	4\\
13.4912	4\\
13.4912	4\\
13.4912	4\\
13.4912	4\\
13.4912	4\\
13.4912	4\\
13.4912	4\\
13.4912	4\\
13.4912	4\\
13.4912	4\\
13.4912	4\\
13.4912	4\\
13.4912	4\\
13.4912	4\\
13.4912	4\\
13.4912	4\\
13.4912	4\\
13.4912	4\\
13.4912	4\\
13.4912	4\\
13.4912	4\\
13.4912	4\\
13.4912	4\\
13.4912	4\\
13.4912	4\\
13.4912	4\\
13.4912	4\\
13.4912	4\\
13.4912	4\\
13.4912	4\\
13.4912	4\\
13.4912	4\\
13.4912	4\\
13.4912	4\\
13.4912	4\\
13.4912	4\\
13.4912	4\\
13.4912	4\\
13.4912	4\\
13.4912	4\\
13.4912	4\\
13.4912	4\\
13.4912	4\\
13.4912	4\\
13.4912	4\\
13.4912	4\\
13.4912	4\\
13.4912	4\\
13.4912	4\\
13.4912	4\\
13.4912	4\\
13.4912	4\\
13.4912	4\\
13.808	3\\
13.808	3\\
13.808	3\\
13.808	3\\
13.808	3\\
13.808	3\\
13.808	3\\
13.808	3\\
13.808	3\\
13.808	3\\
13.808	3\\
13.808	3\\
13.808	3\\
13.808	3\\
13.808	3\\
13.808	3\\
13.808	3\\
13.808	3\\
13.808	3\\
13.808	3\\
13.808	3\\
13.808	3\\
13.808	3\\
13.808	3\\
13.808	3\\
13.808	3\\
13.808	3\\
13.808	3\\
13.808	3\\
13.808	3\\
13.808	3\\
13.808	3\\
13.808	3\\
13.808	3\\
13.808	3\\
13.808	3\\
13.808	3\\
13.808	3\\
13.808	3\\
13.808	3\\
13.808	3\\
13.808	3\\
13.808	3\\
13.808	3\\
13.808	3\\
13.808	3\\
13.808	3\\
13.808	3\\
13.808	3\\
13.808	3\\
13.808	3\\
13.808	3\\
13.808	3\\
13.808	3\\
13.808	3\\
14.04	4\\
14.04	4\\
14.04	4\\
14.04	4\\
14.04	4\\
14.04	4\\
14.04	4\\
14.04	4\\
14.04	4\\
14.04	4\\
14.04	4\\
14.04	4\\
14.04	4\\
14.04	4\\
14.04	4\\
14.04	4\\
14.04	4\\
14.04	4\\
14.04	4\\
14.04	4\\
14.04	4\\
14.04	4\\
14.04	4\\
14.04	4\\
14.04	4\\
14.04	4\\
14.04	4\\
14.04	4\\
14.04	4\\
14.04	4\\
14.04	4\\
14.04	4\\
14.04	4\\
14.04	4\\
14.04	4\\
14.04	4\\
14.04	4\\
14.04	4\\
14.04	4\\
14.04	4\\
14.04	4\\
14.04	4\\
14.04	4\\
14.04	4\\
14.04	4\\
14.04	4\\
14.04	4\\
14.04	4\\
14.04	4\\
14.04	4\\
14.04	4\\
14.04	4\\
14.04	4\\
14.04	4\\
14.04	4\\
14.04	4\\
14.3344	3\\
14.3344	3\\
14.3344	3\\
14.3344	3\\
14.3344	3\\
14.3344	3\\
14.3344	3\\
14.3344	3\\
14.3344	3\\
14.3344	3\\
14.3344	3\\
14.3344	3\\
14.3344	3\\
14.3344	3\\
14.3344	3\\
14.3344	3\\
14.3344	3\\
14.3344	3\\
14.3344	3\\
14.3344	3\\
14.3344	3\\
14.3344	3\\
14.3344	3\\
14.3344	3\\
14.3344	3\\
14.3344	3\\
14.3344	3\\
14.3344	3\\
14.3344	3\\
14.3344	3\\
14.3344	3\\
14.3344	3\\
14.3344	3\\
14.3344	3\\
14.3344	3\\
14.3344	3\\
14.3344	3\\
14.3344	3\\
14.3344	3\\
14.3344	3\\
14.3344	3\\
14.3344	3\\
14.3344	3\\
14.3344	3\\
14.3344	3\\
14.3344	3\\
14.3344	3\\
14.3344	3\\
14.3344	3\\
14.3344	3\\
14.3344	3\\
14.3344	3\\
14.3344	3\\
14.3344	3\\
14.3344	3\\
14.3344	3\\
14.3344	3\\
14.736	2\\
14.736	2\\
14.736	2\\
14.736	2\\
14.736	2\\
14.736	2\\
14.736	2\\
14.736	2\\
14.736	2\\
14.736	2\\
14.736	2\\
14.736	2\\
14.736	2\\
14.736	2\\
14.736	2\\
14.736	2\\
14.736	2\\
14.736	2\\
14.736	2\\
14.736	2\\
14.736	2\\
14.736	2\\
14.736	2\\
14.736	2\\
14.736	2\\
14.736	2\\
14.736	2\\
14.736	2\\
14.736	2\\
14.736	2\\
14.736	2\\
14.736	2\\
14.736	2\\
14.736	2\\
14.736	2\\
14.736	2\\
14.736	2\\
14.736	2\\
14.736	2\\
14.736	2\\
14.736	2\\
14.736	2\\
14.736	2\\
14.736	2\\
14.736	2\\
14.736	2\\
14.736	2\\
14.736	2\\
14.736	2\\
14.736	2\\
14.736	2\\
14.736	2\\
14.736	2\\
14.736	2\\
14.736	2\\
14.736	2\\
14.736	2\\
14.736	2\\
14.8384	4\\
14.8384	4\\
14.8384	4\\
14.8384	4\\
14.8384	4\\
14.8384	4\\
14.8384	4\\
14.8384	4\\
14.8384	4\\
14.8384	4\\
14.8384	4\\
14.8384	4\\
14.8384	4\\
14.8384	4\\
14.8384	4\\
14.8384	4\\
14.8384	4\\
14.8384	4\\
14.8384	4\\
14.8384	4\\
14.8384	4\\
14.8384	4\\
14.8384	4\\
14.8384	4\\
14.8384	4\\
14.8384	4\\
14.8384	4\\
14.8384	4\\
14.8384	4\\
14.8384	4\\
14.8384	4\\
14.8384	4\\
14.8384	4\\
14.8384	4\\
14.8384	4\\
14.8384	4\\
14.8384	4\\
14.8384	4\\
14.8384	4\\
14.8384	4\\
14.8384	4\\
14.8384	4\\
14.8384	4\\
14.8384	4\\
14.8384	4\\
14.8384	4\\
14.8384	4\\
14.8384	4\\
14.8384	4\\
14.8384	4\\
14.8384	4\\
14.8384	4\\
14.8384	4\\
14.8384	4\\
14.8384	4\\
14.8384	4\\
14.8384	4\\
14.8384	4\\
14.8384	4\\
15.1036	3\\
15.1036	3\\
15.1036	3\\
15.1036	3\\
15.1036	3\\
15.1036	3\\
15.1036	3\\
15.1036	3\\
15.1036	3\\
15.1036	3\\
15.1036	3\\
15.1036	3\\
15.1036	3\\
15.1036	3\\
15.1036	3\\
15.1036	3\\
15.1036	3\\
15.1036	3\\
15.1036	3\\
15.1036	3\\
15.1036	3\\
15.1036	3\\
15.1036	3\\
15.1036	3\\
15.1036	3\\
15.1036	3\\
15.1036	3\\
15.1036	3\\
15.1036	3\\
15.1036	3\\
15.1036	3\\
15.1036	3\\
15.1036	3\\
15.1036	3\\
15.1036	3\\
15.1036	3\\
15.1036	3\\
15.1036	3\\
15.1036	3\\
15.1036	3\\
15.1036	3\\
15.1036	3\\
15.1036	3\\
15.1036	3\\
15.1036	3\\
15.1036	3\\
15.1036	3\\
15.1036	3\\
15.1036	3\\
15.1036	3\\
15.1036	3\\
15.1036	3\\
15.1036	3\\
15.1036	3\\
15.1036	3\\
15.1036	3\\
15.1036	3\\
15.1036	3\\
15.1036	3\\
15.1036	3\\
15.3392	4\\
15.3392	4\\
15.3392	4\\
15.3392	4\\
15.3392	4\\
15.3392	4\\
15.3392	4\\
15.3392	4\\
15.3392	4\\
15.3392	4\\
15.3392	4\\
15.3392	4\\
15.3392	4\\
15.3392	4\\
15.3392	4\\
15.3392	4\\
15.3392	4\\
15.3392	4\\
15.3392	4\\
15.3392	4\\
15.3392	4\\
15.3392	4\\
15.3392	4\\
15.3392	4\\
15.3392	4\\
15.3392	4\\
15.3392	4\\
15.3392	4\\
15.3392	4\\
15.3392	4\\
15.3392	4\\
15.3392	4\\
15.3392	4\\
15.3392	4\\
15.3392	4\\
15.3392	4\\
15.3392	4\\
15.3392	4\\
15.3392	4\\
15.3392	4\\
15.3392	4\\
15.3392	4\\
15.3392	4\\
15.3392	4\\
15.3392	4\\
15.3392	4\\
15.3392	4\\
15.3392	4\\
15.3392	4\\
15.3392	4\\
15.3392	4\\
15.3392	4\\
15.3392	4\\
15.3392	4\\
15.3392	4\\
15.3392	4\\
15.3392	4\\
15.3392	4\\
15.3392	4\\
15.3392	4\\
15.3392	4\\
15.5264	3\\
15.5264	3\\
15.5264	3\\
15.5264	3\\
15.5264	3\\
15.5264	3\\
15.5264	3\\
15.5264	3\\
15.5264	3\\
15.5264	3\\
15.5264	3\\
15.5264	3\\
15.5264	3\\
15.5264	3\\
15.5264	3\\
15.5264	3\\
15.5264	3\\
15.5264	3\\
15.5264	3\\
15.5264	3\\
15.5264	3\\
15.5264	3\\
15.5264	3\\
15.5264	3\\
15.5264	3\\
15.5264	3\\
15.5264	3\\
15.5264	3\\
15.5264	3\\
15.5264	3\\
15.5264	3\\
15.5264	3\\
15.5264	3\\
15.5264	3\\
15.5264	3\\
15.5264	3\\
15.5264	3\\
15.5264	3\\
15.5264	3\\
15.5264	3\\
15.5264	3\\
15.5264	3\\
15.5264	3\\
15.5264	3\\
15.5264	3\\
15.5264	3\\
15.5264	3\\
15.5264	3\\
15.5264	3\\
15.5264	3\\
15.5264	3\\
15.5264	3\\
15.5264	3\\
15.5264	3\\
15.5264	3\\
15.5264	3\\
15.5264	3\\
15.5264	3\\
15.5264	3\\
15.5264	3\\
15.5264	3\\
15.5264	3\\
15.9808	4\\
15.9808	4\\
15.9808	4\\
15.9808	4\\
15.9808	4\\
15.9808	4\\
15.9808	4\\
15.9808	4\\
15.9808	4\\
15.9808	4\\
15.9808	4\\
15.9808	4\\
15.9808	4\\
15.9808	4\\
15.9808	4\\
15.9808	4\\
15.9808	4\\
15.9808	4\\
15.9808	4\\
15.9808	4\\
15.9808	4\\
15.9808	4\\
15.9808	4\\
15.9808	4\\
15.9808	4\\
15.9808	4\\
15.9808	4\\
15.9808	4\\
15.9808	4\\
15.9808	4\\
15.9808	4\\
15.9808	4\\
15.9808	4\\
15.9808	4\\
15.9808	4\\
15.9808	4\\
15.9808	4\\
15.9808	4\\
15.9808	4\\
15.9808	4\\
15.9808	4\\
15.9808	4\\
15.9808	4\\
15.9808	4\\
15.9808	4\\
15.9808	4\\
15.9808	4\\
15.9808	4\\
15.9808	4\\
15.9808	4\\
15.9808	4\\
15.9808	4\\
15.9808	4\\
15.9808	4\\
15.9808	4\\
15.9808	4\\
15.9808	4\\
15.9808	4\\
15.9808	4\\
15.9808	4\\
15.9808	4\\
15.9808	4\\
15.9808	4\\
};
\label{plotyyref:leg1}

\end{axis}

\begin{axis}[%
width=\fwidth,
height=0.774\fheight,
at={(0\fwidth,0\fheight)},
scale only axis,
xmin=12,
xmax=16,
every outer y axis line/.append style={mycolor2},
every y tick label/.append style={font=\color{mycolor2}\tiny},
every y tick/.append style={mycolor2},
ymin=0,
ymax=600,
ytick={  0, 200, 400, 600},
axis x line*=bottom,
axis y line*=right,
legend style={legend cell align=left, align=left, draw=white!15!black},
ymajorgrids
]
\addlegendimage{/pgfplots/refstyle=plotyyref:leg1}
\addlegendentry{CellId}
\addplot [color=mycolor2, line width=0.8pt]
  table[row sep=crcr]{%
12.05	75.21344\\
12.1	75.21344\\
12.15	149.42304\\
12.2	81.2856\\
12.25	170.84032\\
12.3	172.18144\\
12.35	79.2608\\
12.4	76.23328\\
12.45	133.57088\\
12.5	256.00352\\
12.55	286.35072\\
12.6	284.49568\\
12.65	397.82176\\
12.7	402.71232\\
12.75	346.5552\\
12.8	343.85696\\
12.85	211.47456\\
12.9000000000001	243.86144\\
12.9500000000001	259.87424\\
13.0000000000001	256.16416\\
13.0500000000001	264.93344\\
13.1000000000001	374.3888\\
13.1500000000001	80.112\\
13.2000000000001	76.05664\\
13.2500000000001	379.11072\\
13.3000000000001	82.97888\\
13.3500000000001	298.66944\\
13.4000000000001	282.30336\\
13.4500000000001	231.71136\\
13.5000000000001	205.24288\\
13.5500000000001	115.34976\\
13.6000000000001	83.30816\\
13.6500000000001	123.61312\\
13.7000000000001	123.61312\\
13.7500000000001	174.37376\\
13.8000000000001	174.37376\\
13.8500000000001	339.64896\\
13.9000000000001	338.9664\\
13.9500000000001	211.30592\\
14.0000000000001	220.0752\\
14.0500000000001	232.56256\\
14.1000000000001	270.49856\\
14.1500000000001	206.24672\\
14.2000000000001	181.11936\\
14.2500000000001	238.79424\\
14.3000000000001	219.73792\\
14.3500000000001	197.14816\\
14.4000000000001	312.32128\\
14.4500000000001	287.69984\\
14.5000000000001	280.27968\\
14.5500000000001	309.4544\\
14.6000000000001	289.89216\\
14.6500000000001	281.6288\\
14.7000000000001	318.05504\\
14.7500000000001	123.62112\\
14.8000000000001	2.19232\\
14.8500000000001	165.61248\\
14.9000000000001	231.20544\\
14.9500000000001	214.1728\\
15.0000000000001	214.34144\\
15.0500000000001	251.10496\\
15.1000000000001	235.75872\\
15.1500000000001	366.12544\\
15.2000000000001	444.02912\\
15.2500000000001	338.29184\\
15.3000000000001	321.93376\\
15.3500000000001	154.81952\\
15.4000000000001	300.34784\\
15.4500000000001	124.7936\\
15.5000000000001	123.44448\\
15.5500000000001	166.9616\\
15.6000000000001	257.176\\
15.6500000000001	309.96032\\
15.7000000000001	319.40416\\
15.7500000000001	205.40352\\
15.8000000000001	183.31168\\
15.8500000000001	213.49824\\
15.9000000000001	304.05792\\
15.9500000000001	213.49824\\
};
\addlegendentry{Throughput [Mbits]}

\end{axis}
\end{tikzpicture}%

%% file: figures/latency_10MB_cornerCase_20us.tex
%
%
\definecolor{mycolor3}{rgb}{0.00000,0.70000,0.70000}%
\definecolor{mycolor2}{rgb}{0.97630,0.98310,0.05380}%
\definecolor{mycolor1}{rgb}{0.2196,0.1725,0.5020}%
\begin{tikzpicture}
\pgfplotsset{every tick label/.append style={font=\tiny}}
\pgfplotsset{scaled y ticks=false}

\begin{axis}[%
ybar=100pt,
width=\fwidth,
height=\fheight,
at={(0\fwidth,0\fheight)},
scale only axis,
bar shift auto,
xtick=data,
enlarge x limits=0.2,
bar width=8pt,
xticklabels={{1.6},{12.8},{25.6}},
xlabel style={font=\scriptsize\color{white!15!black}},
xlabel={Delay D [ms]},
ymin=0,
ymax=0.016,
ylabel style={font=\scriptsize\color{white!15!black}},
ylabel={Latency [s]},
axis background/.style={fill=white},
xmajorgrids,
ymajorgrids,
ylabel shift = -5 pt,
xlabel shift = -5 pt,
yticklabel shift = -2 pt,
xticklabel shift = -5 pt,
yticklabel style={
            /pgf/number format/fixed,
            /pgf/number format/precision=3,
            /pgf/number format/fixed zerofill
        },
        scaled y ticks=false,
legend style={font=\scriptsize,at={(0.99,0.5)}, anchor=east, legend cell align=left, align=left, draw=white!15!black}
]


\addplot[fill=mycolor1] coordinates {(1,0.00973376700345201) (2,0.00702147879040495) (3,0.0149143252249734)};
\addplot[fill=mycolor2] coordinates {(1,0.00160106520118727) (2,0.00209109787579124) (3,0.00273510037943536)};
\legend{Fixed TTT, Dynamic TTT}



\end{axis}
\end{tikzpicture}%